\documentclass[preprintnumbers,prb,superscriptaddress,preprint]{revtex4-1}

\usepackage{amsmath}
\usepackage{graphicx}
\usepackage{dcolumn}
\usepackage{latexsym,bm}
\usepackage{hyperref}
\usepackage{amsfonts}
\usepackage{tabularx}
\usepackage{cases}
\usepackage{amssymb}
\usepackage{enumerate}
\usepackage{mathtools}
\usepackage{etoolbox} % for \appto
\usepackage{lipsum} % for mock text
\usepackage[capitalize]{cleveref}
\usepackage{ulem}
\usepackage{xcolor}

\begin{document}

\title{Topological Phase Transition Induced by Image Potential States in MXenes: A Theoretical Investigation}

\author{Mengying Wang}
\affiliation{Department of Physics, Shanghai Normal University, Shanghai 200234, China}

\author{Mohammad Khazaei}
\affiliation{Department of Physics, Yokohama National University, Yokohama 240-8501, Japan}

\author{Yoshiyuki Kawazoe}
\affiliation{New Industry Creation Hatchery Center, Tohoku University, Sendai, 980-8579, Japan}
\affiliation{School of Physics, Institute of Science and Center of Excellence in Advanced Functional Materials, Suranaree University of Technology, Nakhon Ratchasima 30000, Thailand}

\author{Yunye Liang}
\email{E-mail: liangyunye@shnu.edu.cn}
\affiliation{Department of Physics, Shanghai Normal University, Shanghai 200234, China}

\date{\today}% It is always \today, today,

\widetext

\begin{abstract}

MXenes, a family of two-dimensional transition metal carbides and nitrides, have various tunable physical and chemical properties. Their diverse prospective applications in electronics and energy storage devices have triggered great interests in science and technology. MXenes can be functionalized by different surface terminations. Some O and F functionalized MXenes monolayers have been predicted to be topological insulators (TIs). However, the reported OH functionalized MXenes TIs are very few and their electronic structures need to be investigated in more detail. It has been revealed that the work functions of MXenes are reduced significantly by OH termination and the image potential (IP) states move close to the Fermi level. The wave functions of these IP states are spatially extensive outside the surfaces. By stacking the OH-functionalized MXenes, the energies of the IP states can be modulated by the interlayer distances of multilayers, because the overlap and hybridization of the wave functions between the neighboring layers are significant. Therefore, these stacking layers are interacted and coupled with IP states. Here, based on first-principles calculations, we demonstrate that the stacking of two-dimensional topologically trivial OH-functionalized MXenes, such as V$_2$HfC$_2$(OH)$_2$, possibly gives rise to the topologically nontrivial energy bands. In other words, the topological properties of V$_2$HfC$_2$(OH)$_2$ multilayers can be modulated by its interlayer distance. An energy band inversion involving IP states is proposed. We expect that these results can advance the future application of MXenes or other low work function multilayer materials as controllable TI devices.
\end{abstract}
\bigskip

\maketitle
\section{Introduction}
MXenes have attracted more and more attention since they were obtained from the chemical exfoliation of MAX phase ceramic compounds.~\cite{M. Naguib,M. Naguib2,M. Naguib3} Their various prospective properties and potential applications have been studied and proposed.~\cite{M. Khazaei, M. Khazaei2, E.Balci2018, M.Khazaei2017, M.Khazaei2019, B.Anasori2017_1, B.M.Jun2018, J.Pang2019, A.L.Ivanovskii2013, N.K.Chaudhari2017, J.Zhu2017, H.Wang2018, X.Li2018, X.Zhang2018, Y.Zhang2018, K.Hantanasirisakul2018, H.Lin2018, C.Zhan2019, J. Zhou, Hongming Weng2015, Yunye_Liang, Khazaei2016,C. Si,L. Li,Z-Q_Huang, A. Champagne} MXenes have a general chemical formula of M$_{n+1}$X$_n$, where M stands for the transition metal, X is C or N, and $n$ = 1$-$3.~\cite{M. Naguib} Up to now, many MXenes, including Ti$_2$C, V$_2$C, Nb$_2$C, Mo$_2$C, Ti$_3$CN, Ti$_3$C$_2$, Ta$_4$C$_3$, and Ti$_3$C$_2$ have been experimentally fabricated. Recently, the family of MXenes has been expanded to ordered double transition metal carbides M$'_2$M$''$C$_2$ and M$'_2$M$''_2$C$_3$, where M$'$ and M$''$ stand for different transition metals.~\cite{B. Anasori,B. Anasori2,B. Anasori3} Determined by the synthesis process, fluorine (F), oxygen (O) or hydroxyl (OH) groups terminate and functionalize the MXenes surfaces. As a result, their physical and chemical properties can be designed.~\cite{M. Khazaei, M. Khazaei2}

In O or F-terminated MXenes, theoretical calculations find that the energy bands around the Fermi level are dominated by $d$-orbitals of the transition metals~\cite{J. Zhou,Yunye_Liang}. Five-fold degeneracy of $d$-orbitals is lifted by the crystal fields of hexagonal lattices. Without considering the spin orbit coupling (SOC), the functionalized MXenes, such as M$_2$CO$_2$ (M=W, Mo) and M$'_2$M$''$C$_2$O$_2$ (M$'$= Mo, W; M$''$= Ti, Zr, W) are zero-gap semiconductors while the valence and the conduction bands touch only at the $\Gamma$-point.\cite{Hongming Weng2015, Khazaei2016, C. Si, L. Li} However, the MXenes containing heavy $4d$ and $5d$ transition metals exhibit significant relativistic SOC. Upon considering the SOC, the degeneracy of the bands at $\Gamma$-point is lifted.\cite{Hongming Weng2015, Khazaei2016, C. Si, L. Li} The band gaps are open, coinciding with the energy bands inversion. These MXene monolayers are demonstrated to be topological insulators (TIs) by non-zero topological $\mathbb{Z}_2$ invariants and with the conducting helical edge states. A variety of MXenes have been theoretically predicted to be TIs, including MXenes carbides or nitrides.\cite{Hongming Weng2015, Yunye_Liang, Khazaei2016,C. Si,L. Li,Z-Q_Huang} However, the reported OH-terminated MXenes TIs are very few and their energy band inversion mechanism needs to be investigated in more detail.   

Based on density functional theory (DFT) calculations, we have previously revealed that OH-functionalized MXenes have ultra-low work functions.\cite{OH} As a result, the image potential (IP) states approach to the Fermi level.\cite{NFE} These facts make the energy bands of OH-functionalized MXenes different from O or F-functionalized MXenes, whose IP states are several electron-volts away from the Fermi levels.\cite{NFE} Therefore, IP states can play important roles in determining the electronic structures of OH-functionalized MXenes.~\cite{J. Zhou} It is noteworthy that the IP states have also been observed in a variety of low-dimensional materials such as graphene and BN nanotubes. The IP states have several distinct features from the bands that come from the atomic orbitals.~\cite{J. Zhao, J. Zhao2, M. Feng, M. Feng2, S. Bose, V. M. Silkin, V. M. Silkin, N. T. Cuong, K. H. Khoo, M. Ishigami, J. Zhao charging, E. R. Margine, E. Balci} For example, the valence band of Sc$_2$C(OH)$_2$ comes from the IP states.~\cite{NFE,J. Zhou} Its wave function $\phi(\vec{r})$ is spatially extensive in the direction normal to the surfaces and the charge density $|\phi(\vec{r})|^2$ reaches its maximum in the vacuum regions.~\cite{J. Zhou} In Sc$_2$C(OH)$_2$ multilayers, the overlap and hybridization of the wave function long tails from the neighboring layers are significant, even when their interlayer distances are relatively large. The energies of IP states can be shifted by the interlayer distances.~\cite{J. Zhou}

Given these facts, we study the electronic structures of the ordered double transition metal carbides M$'_2$M$''$C$_2$(OH)$_2$, where M$'$ represents V, Nb, Ta and M$''$ is Ti, Zr, Hf. On the basis of first-principles calculations, we address that the IP states in these MXenes are crucial because they appear near the Fermi levels. We demonstrate that their topological properties are related to the IP states and can be modulated by the interlayer distances. By changing the interlayer distances, the IP energy bands can be shifted. The band alignment is changed and the energy band inversion is induced resulting in topologically nontrivial bands. For instance, in V$_2$HfC$_2$(OH)$_2$ multilayers, if the interlayer distance changes in the range from {30~\AA} to 20~\AA, the trivial material becomes nontrivial strong topological one. 

\section{Calculation methods}
\label{Calculation_method}

First-principles calculations were performed within the framework of DFT by the Vienna \textit{ab initio} simulation package (VASP).~\cite{G. Kress, G. Kress2, G. Kress3, G. Kress4, G. Kress5} In the calculations, the exchange-correlation functional proposed by Perdew, Burke and Ernzerhof (GGA-PBE) was adopted with the projector augmented wave (PAW) method.~\cite{J. P. Perdew, G. Kress2} A cutoff energy of 520~eV was used for the plane wave basis sets.~\cite{M. Khazaei} All the atoms were fully relaxed until the force acting on each atom was less than 10\textsuperscript{-4}~eV/{\AA}. The convergence of the energy was less than 10\textsuperscript{-8}~eV. A 21$\times$21$\times$3 Monkhorst-Pack $k$-mesh was used to perform the geometrical relaxation calculations and a 33$\times$33$\times$3 $k$-mesh was employed for all energy band calculations.~\cite{H. J. Monkhorst, M. Methfessel} 

Since GGA–PBE functional is valid only for the short range and the predicted band gap width is typically underestimated.~\cite{J. Perdew} To increase the accuracy and improve the results, the Heyd–Scuseria–Ernzerhof (HSE06) screened hybrid functional is performed.~\cite{T. M. Henderson, M. Ernzerhof, C. Adamo} One-fourth of the PBE exchange is replaced by the Hartree-Fock exact exchange, and the full PBE correlation energy is included. The van der Waals (vdW) interaction between the neighboring layers is also taken into consideration, and the empirical correction method presented by Grimme (DFT-D2), which has been proven reliable for describing in long range, was adopted.~\cite{S. Grimme} These results are presented in the Supplemental Material.

As far as we know, the successful synthesis of ordered M$'_2$M$''$C$_2$(OH)$_2$ MXenes have not been achieved yet. The phonon spectra were calculated by Phonopy to confirm their local stabilities ~\cite{A.Togo,K. Parlinski}. The force constants were extracted from the 4$\times$4$\times$1 supercells.~\cite{K. Parlinski} 

The understanding of the energy bands on the footing of symmetry is always helpful, and Irvsp was applied to analyze the irreducible representations.~\cite{irvsp} To confirm the topological properties, the effective tight-binding Hamiltonian was extracted by Wannier90, and the analysis was realized by WannierTools.~\cite{wannier90,wanniertools}

\section{Results and Discussion}
\label{Results and Discussion}
\subsection{Structural Information}
\label{Geometries}

Hereafter, V$_2$HfC$_2$(OH)$_2$ is used as the representative of M$'_2$M$''$C$_2$(OH)$_2$ because the energy bands of other materials are very similar to that of V$_2$HfC$_2$(OH)$_2$. When OH groups are absorbed on the surfaces, there are three possible sites as shown in FIG.~\ref{structure_OH}, and they are named as A, B, and T.~\cite{J. Zhou} Site T is on the top of the outmost V atom; site B is the hollow site above the C atom; and site A is the hollow site above the Hf atom of the same surface. Regarding the combinations of these three absorption sites on two surfaces, six distinct configurations can be constructed.~\cite{M. Naguib, M. Khazaei} All these possible structures are fully relaxed and their energies are summarized in TABLE.~\ref{stablestructure}. The calculations reveal that BB models are the most energy favorable structures after relaxations.

Although materials such as ScCx(OH), and ordered double transition metals MXenes exist experimentally, the successful synthesis of M$'_2$M$''$C$_2$(OH)$_2$ have not been achieved yet.~\cite{J.Zhou2019,S.J.Hwu1986} Therefore, the stabilities of these energy favorable structures are investigated by phonon calculations in FIG. \ref{phonon}. They are dynamically stable because their vibrational frequencies are all positive. In the phonon calculations, the vacuum sizes are 30~{\AA}. The phonon results with smaller vacuum sizes (20~{\AA}) were also investigated, shown in FIG. S1 in the Supplemental Material. No significant changes were observed. For the sake of visibility of the acoustic bands, two highest O-H stretching bands are not shown in the phonon spectra figures. 
 
\begin{table}[tb]
\caption{
The relative energies (in $eV$ per unit cell) of six possible M$'_2$M$''$C$_2$(OH)$_2$ configurations, where M$'$=V, Nb, Ta and M$''$=Ti, Zr, Hf. The adsorption sites of A, B, T are indicated in FIG.~\ref{structure_OH}. The energy of the most favorable structure is set to zero and the energies of the other structures are relative to it. 
}\label{stablestructure}
\begin{tabular*}{0.5\textwidth}{@{\extracolsep{\fill}}c|cccccc}
 	\hline
 	\hline
sites of OH & BB & BA & AA & TA & TB & TT \\
         \hline
 V$_2$TiC$_2$(OH)$_2$ & 0.00 & 0.24 & 0.31 & 0.79 & 0.90 & 1.32 \\
 V$_2$ZrC$_2$(OH)$_2$ & 0.00 & 0.13 & 0.14 & 0.45 & 0.68 & 0.90 \\
 V$_2$HfC$_2$(OH)$_2$ & 0.00 & 0.17 & 0.19 & 0.57 & 0.77 & 1.03 \\
Nb$_2$TiC$_2$(OH)$_2$ & 0.00 & 0.44 & 0.68 & 1.14 & 1.04 & 1.77 \\
Nb$_2$ZrC$_2$(OH)$_2$ & 0.00 & 0.43 & 0.62 & 1.03 & 0.97 & 1.56 \\
Nb$_2$HfC$_2$(OH)$_2$ & 0.00 & 0.48 & 0.75 & 1.18 & 1.06 & 1.79 \\
Ta$_2$TiC$_2$(OH)$_2$  & 0.00 & 0.46 & 0.61 & 0.97 & 0.89 & 1.48 \\
Ta$_2$ZrC$_2$(OH)$_2$ & 0.00 & 0.42 & 0.55 & 0.85 & 0.82 & 1.32  \\
Ta$_2$HfC$_2$(OH)$_2$ & 0.00 & 0.49 & 0.70 & 1.02 & 0.91 & 1.51 \\
 	\hline
 	\hline
\end{tabular*}
\end{table}

In FIG.\ref{structure_OH}, V$_2$HfC$_2$(OH)$_2$ has a hexagonal cell, whose lattice constant $a$ is about 3.01~{\AA}. Its space group belongs to $P\overline{3}m1$. The Hf atom is the center of the primitive cell. Its Wyckoff position is 1$a$, whose site symmetry is $D_{3d}$. Due to the crystal field, its five-fold $d$-orbitals split into three groups. The $d_{x^2-y^2}$ and $d_{xy}$ orbitals form two-dimensional $E_g$ representation. The $d_{xz}$ and $d_{yz}$ orbitals form another $E_g$ representation and $d_{z^2}$ orbital forms one-dimensional $A_{1g}$ representation. The site symmetries of V atoms are $C_{3v}$ and the $d$-orbitals are categorized into three sets: ($d_{x^2-y^2}$, $d_{xy}$) and ($d_{xz}$, $d_{yz}$) with two-fold degeneracy belonging to $E$ representations and $d_{z^2}$ forms the one-dimensional $A_1$ representation. 

\subsection{Electronic Properties}
\label{Electronic Properties}

Only the most energy favorable BB-structure of V$_2$HfC$_2$(OH)$_2$ is considered and the fat-bands are plotted in FIG.~\ref{fatband_V2Hf_30}. The wave functions are respectively projected onto $s$, $p$, and $d$-orbitals of all atoms in the primitive cell and the sizes of the dots are proportional to the projection weights. It is found that the conduction and valence bands are separated by a direct tiny gap. V$_2$HfC$_2$(OH)$_2$ is a semimetal, because the bottom of the conduction band at $\Gamma$-point and the top of valence band at M-point have a very small overlap. In FIG.~\ref{fatband_V2Hf_30}(a), the weights from $s$-orbitals are very small, and in FIG.~\ref{fatband_V2Hf_30}(b), $p$-orbitals have the contribution to the top of the valence band at M-point. As shown in FIG.~\ref{fatband_V2Hf_30}(c),  $d$-orbitals are overwhelmed and determinant. However, around the $\Gamma$-point, the contributions of $d$-orbital to the conduction and valence bands become remarkably small. It means that these states are not ascribed to any atomic orbitals and they mainly come from the IP states.

In the previous studies, it is revealed that the IP states of the OH-terminated MXenes are close to the Fermi level.~\cite{NFE} IP states do not come from any atomic orbitals. Hence, the projection weights of the IP states onto atomic orbitals are small. The energy bands of IP states possess parabolic energy dispersions near the $\Gamma$-point, which are analogous to the free electron states.~\cite{NFE,M. Feng, M. Feng2} In most 2D materials, such as graphene and BN, the energies of IP states are several electron-volts away from the Fermi levels.~\cite{J. Zhao, J. Zhao2, M. Feng, M. Feng2, S. Bose, V. M. Silkin, V. M. Silkin, N. T. Cuong, K. H. Khoo, M. Ishigami, J. Zhao charging, E. R. Margine, E. Balci} For this reason, they play less important roles in the determination of the physical or chemical properties and thus are always neglected. 

In OH-functionalized MXene multilayers, such as Sc$_2$C(OH)$_2$, the wave functions of the IP states are spatially extensive and their energetic positions are dependent on the interlayer distances.~\cite{J. Zhou} By tuning the interlayer distances, these energy bands shift significantly due to the overlap and hybridization of the wave function tails from the neighboring layers, while other bands coming from the atomic orbitals have no visible changes.~\cite{J. Zhou} To investigate the spatial distributions of the wave functions in V$_2$HfC$_2$(OH)$_2$, the decomposed charge densities of valence and conduction bands at $\Gamma$-point are plotted and averaged along the $c$-axis in FIG.~\ref{chgden_V2Hf}. The corresponding side views are shown below them. In FIG.~\ref{chgden_V2Hf}(a), the interlayer distance is 30~{\AA}. These wave functions have very long tails. The wave functions are very extensive and decay to zero gradually. The charge density maxima are 1.2~{\AA} away from the outmost H atoms. Since V$_2$HfC$_2$(OH)$_2$ possess inversion symmetry, the energy bands can be classified by their parities. For the wave functions with odd parities, they must be zero in the central Hf atom. FIG.~\ref{chgden_V2Hf}(a) indicates that the valence band has odd parity because its wave function density is zero around the central Hf atom but the conduction band is even due to its nonzero peak in the center. These results are consistent with the irreducible representations of the conduction and valence bands.~\cite{irvsp} As given in TABLE.~\ref{30}, the valence (conduction) band belongs to $A_{2u}$ ($A_{1g}$) representation with odd (even) parity. Along the $\Sigma$-axis (which connects $\Gamma$ and M-points), the symmetry is reduced to $C_s$. Both conduction and valence bands belong to $A'$ representation. Along $\Lambda$-axis (which connects $\Gamma$ and K-points), the symmetry becomes $C_2$. The representations of valence and conduction bands are $B$ and $A$, respectively.

\begin{table}[tb]
\caption{
The irreducible representations of valence and conduction bands along the high symmetrical axes or points in V$_2$HfC$_2$(OH)$_2$ multilayer with the interlayer distance $l=$30~{\AA}. $\Sigma$-axis, whose symmetrical operator is $C_s$, connects M and $\Gamma$-points. $\Lambda$-axis, whose symmetrical operator is $C_2$, connects K and $\Gamma$-points. The symmetrical operator of $\Gamma$-point is $D_{3d}$. Only the points close to $\Gamma$-point are considered. The high symmetrical points $\Gamma$, K and M are shown in FIG.~\ref{structure_OH}(c).
}\label{30}
\begin{tabular*}{0.5\textwidth}{@{\extracolsep{\fill}}c|ccc}
 	\hline
 	\hline
  $l=30$~\AA  & $\Sigma(C_s)$ & $\Gamma(D_{3d})$ & $\Lambda(C_2)$  \\
         \hline
 valence band & $A'$ & $A_{2u}$ & $B$  \\
 conduction band& $A'$ & $A_{1g}$ & $A$ \\
 	\hline
 	\hline
\end{tabular*}
\end{table}

In the calculations, the periodical boundary condition is applied. Therefore, the interactions from the neighboring periodical images may be crucial. FIG.~\ref{chgden_V2Hf}(a) indicates that V$_2$HfC$_2$(OH)$_2$ is well separated when the interlayer distance is 30~\AA, because the overlap of wave function tails from the neighboring layers is very small. Therefore, its energy bands represent the results of a V$_2$HfC$_2$(OH)$_2$ monolayer. Experimentally, MXenes multilayers are fabricated by the stacking of monolayers and the interlayer distances can be modulated in a wide range~\cite{Y. Ma}. As a comparison, FIG.~\ref{chgden_V2Hf}(b) shows the decomposed charge densities of V$_2$HfC$_2$(OH)$_2$ multilayer with the interlayer distances of 20~{\AA}. The differences are noticeable. By reducing the interlayer distances from 30 to 20~{\AA}, the overlap of the tails and their hybridization induce two significant differences: the growing nonzero valence band tails in the vacuum region as indicated by the arrows in FIG.~\ref{chgden_V2Hf}(b) and the change of the parities of valence and the conduction bands. 

\begin{table}[tb]
\caption{
The irreducible representations of valence and conduction bands along high symmetrical axes or points in V$_2$HfC$_2$(OH)$_2$ multilayer with the interlayer distance $l=$20~{\AA}.
}\label{20}
\begin{tabular*}{0.5\textwidth}{@{\extracolsep{\fill}}c|ccc}
 	\hline
 	\hline
  $l=20$~\AA  & $\Sigma(C_s)$ & $\Gamma(D_{3d})$ & $\Lambda(C_2)$  \\
         \hline
 valence band & $A'$ & $A_{1g}$ & $A$  \\
 conduction band& $A'$ & $A_{2u}$ & $B$ \\
 	\hline
 	\hline
\end{tabular*}
\end{table}

TABLE.~\ref{20} is the irreducible representations of the valence and conduction bands around $\Gamma$-point when the interlayer distance is reduced to 20~{\AA}. Compared with the previous results in TABLE.~\ref{30}, the representation of the valence band at $\Gamma$-point changes from $A_{2u}$ into $A_{1g}$, while the conduction band changes from $A_{1g}$ into $A_{2u}$ representation. In FIG.~\ref{chgden_V2Hf}(b), the valence band has a charge density peak in the center of the cell while the charge density of the conduction band is zero. Along $\Lambda$-axis, the exchange of the irreducible representations between the valence and the conduction bands can be also confirmed in TABLE.~\ref{20}. 

To understand what happens to the electronic structures when the interlayer distances are tuned, the energy bands of V$_2$HfC$_2$(OH)$_2$ with different interlayer distances are plotted in FIG.~\ref{band_distance}. It is observed that by reducing the interlayer distances, the energy band gaps at $\Gamma$-point become much smaller, and when this distance is about 25~{\AA}, the band gap is too small to be found. By further reducing the interlayer distance, the valence band and the conduction band interchange their roles. At $\Gamma$-point, the band with even parity ($A_{1g}$) becomes the valence band, however, the band with odd parity ($A_{2u}$) is the conduction band. Along $\Sigma$-axis ($\Gamma$-M direction), both bands have the same $A'$ representations, the anti-crossing of them keeps the gap open. Along $\Lambda$-axis ($\Gamma$-K direction), two bands belong to different representations and they are crossing. Therefore, the Dirac cone can be found along $\Lambda$-axis. As shown in the inset of FIG.~\ref{band_distance}(c), in the presence of SOC, the closed energy band gap is opened again. In FIG.~\ref{fatband_V2Hf_30}, the heavy elements contribute less weights to wave functions of these bands, therefore the width of the energy band gap is very tiny ($\sim$0.0017~eV).

When the interlayer distance is about 20~{\AA}, the wave function tails of the neighboring layers are interacted and coupled along $c$-axis resulting in significant overlaps. The energy dispersions in this direction should be taken into consideration and are presented in the upper panel of FIG.~\ref{3dband}. As a comparison, the energy bands of the structure with 30~{\AA} vacuum space is also plotted along the same high symmetrical axes. It is seen that mostly their energy bands are very similar. The main changes are indicated by the arrows. By reducing the interlayer distances,  the energy differences between conduction and valence bands along $\Gamma-$A axis become visible. The energy bands evolve in the same way upon including the vdW interaction or adopting the HSE06 hybrid functional in the calculations.~\cite{T. M. Henderson, M. Ernzerhof, C. Adamo, S. Grimme}  These results are given in FIG. S2 and  FIG. S3 of the Supplemental Material. The energy bands of other multilayer structures with the changes of interlayer distances are shown in the Supplemental Material (FIG. S4-FIG. S11). 

These changes result from the wave functions hybridizations of IP states. As it has been revealed, the wave functions of the IP states along the $c$-axis are analogous to the radial part of H atomic wave functions.~\cite{J. Zhou} Therefore, the decrease of the interlayer distances of V$_2$HfC$_2$(OH)$_2$ is similar to the approaching of two H atoms. In the H$_2$ molecule, the bonding state has the symmetrical combination ($\phi_1+\phi_2$) of the atomic orbitals and lower energy, while the anti-symmetrical combination ($\phi_1-\phi_2$) gives the anti-bonding state and higher energy. The similar processes happen and are shown in FIG.~\ref{band_inv}. For simplicity, only $\Gamma$ and A-points have been involved, because the topological properties are related to these points.~\cite{Liang Fu} 

Suppose the $j$th stacking monolayer provides two IP states ($\phi^-_j$ and $\phi^+_j$), whose parities are distinct. Since spatial inversion symmetry is the symmetrical operation of $\Gamma$ and A-points, the hybridization of the wave functions with distinct parities is impossible. Thus, adopting tight-binding approximation, the Bloch wave functions are:
\begin{gather*}
\psi^{\pm}_k=\frac{1}{\sqrt{N}}\sum_je^{i k R_j}\phi^{\pm}_j  
\label{tightbinding}
\end{gather*}
where $R_j$ is the coordinate of the $j$th monolayer center, $k$ is the wave vector along $\Gamma$-A axis and $N$ is the total number of layers. The wave functions of $\Gamma$ and A-points becomes:
\begin{align}
\psi^{\pm}_{\Gamma}=\frac{1}{\sqrt{N}}\sum_j(\phi^{\pm}_{2j}+\phi^{\pm}_{2j+1}) \\
\psi^{\pm}_A=\frac{1}{\sqrt{N}}\sum_j(\phi^{\pm}_{2j}-\phi^{\pm}_{2j+1})   \label{A-point}
\end{align}

When these monolayers are well separated by 30~{\AA} thickness, FIG.~\ref{band_distance}(a) indicates that the energy of the odd state ($\phi^-_j$) is lower than that of even one ($\phi^+_j$). If the interlayer distance is decreasing, the overlap of the wave function tails from the neighboring layers becomes much more significant. The evolution of the wave functions at $\Gamma$-point with respective to the interlayer distances is sketched in FIG.~\ref{band_inv}(a) . The neighboring even parity states (blue) have a symmetrical combination, which are similar to the bonding states of H$_2$ molecule. The energies get lower when the neighboring layers are closer, because the symmetrical combination makes the wave function much more delocalized. Its energy level shifts downwards. The combination of the neighboring odd parity states (red) is anti-symmetrical and the energy level moves upwards. These two energy levels interchange their positions when the interlayer distance is about 25~{\AA} in FIG.~\ref{band_distance}(b). The situation is reversed at A-point, simply because the neighboring states are anti-phase in Eq.~\ref{A-point}. The decreasing of the interlayer distance delocalizes the wave function of the odd parity state (red), while the even state becomes localized (blue). Their energy levels shift in opposite ways, and their energy difference is much greater. This is consistent with the results shown in FIG.~\ref{3dband}. 

The topology of energy bands of layered materials will be tuned when they are stacked.~\cite{D. Wang, H. Weng_PRX} The realignment of the energy bands and the interchange of the parities trigger the energy band inversion, which plays important role in the topological transition of energy bands. These facts inspire us to explore the topological properties of V$_2$HfC$_2$(OH)$_2$ multilayers.

\subsection{Topological Properties}
\label{Energy Band Inversion}

In 3D, there are four $\mathbb{Z}_2$ topological invariants written as $\nu_0$;($\nu_1\nu_2\nu_3$).~\cite{Liang Fu} With spatial inversion symmetry, the topological invariants can be simply evaluated from the parities of all occupied bands at the time-reversal invariant momentum (TRIM) points.~\cite{Liang Fu} In V$_2$HfC$_2$(OH)$_2$ hexagonal lattice, the TRIM points $\vec{k}_i$ include $\Gamma$, A, three M and three L-points. By investigating the products of parities of all occupied bands at these TRIM points, $\delta(\vec{k}_i)=\prod_{n=1}^N\xi_{n} (\vec{k}_i)$, the results are tabulated in TABLE.~\ref{delta}, where $\vec{k}_i$ is TRIM point, $\xi_n(\vec{k}_i)$ is $+1$ or $-1$ for even or odd parity of the $n$th occupied band (not including the Kramers degenerate partner) and $N$ is the number of occupied bands (counting only one of the Kramers degenerate pairs). Thus when the interlayer distance is large than 30, V$_2$HfC$_2$(OH)$_2$ is topologically trivial because its $\mathbb{Z}_2$ invariants are 0;(0,0,0). By reducing the interlayer distances ($l$=20~\AA), it becomes 3D strong TI whose $\mathbb{Z}_2$ invariants are 1;(0,0,0).

\begin{table}[tb]
\caption{
$\delta(\vec{k}_i)$ represents the products of the parities from all occupied energy bands at TRIM points $\vec{k}_i$ ($\Gamma$, A, three M and three L-points). The positions of these TRIM points are shown in FIG.~\ref{structure_OH}(c). $l$ is the interlayer distance.
}\label{delta}
\begin{tabular*}{0.5\textwidth}{@{\extracolsep{\fill}}c|cccc}
 	\hline
 	\hline
    & $\delta$($\Gamma$) & $\delta$(A) & $\delta$(M) & $\delta$(L)  \\
         \hline
 $l=30$~\AA& 1 & 1 & 1  &  1 \\
 $l=20$~\AA& -1 & 1 & 1  &  1 \\
 	\hline
 	\hline
\end{tabular*}
\end{table}

One remarkable feature of the strong 3D TI is the odd number of Dirac-cone-type boundary states, which are robust due to the protection of time-reversal symmetry. In order to further confirm the nontrivial topological properties ($l$=20~{\AA}), these boundary states were explored. In the calculations, the effective tight-binding Hamiltonian is extracted by applying Wannier functions. Merely $d$-orbitals of V and Hf atoms, $p$-orbitals of C and O atoms are considered, because the energy bands near the Fermi energy are mainly stem from these orbitals. FIG.~\ref{edge_state}(a) shows the (0001) surface states and a single Dirac cone is observed around $\Gamma$-point. The boundary states along $y$-axis, which is defined in FIG.~\ref{structure_OH}(a), is found in FIG.~\ref{edge_state}(b). Therefore, the energy bands of V$_2$HfC$_2$(OH)$_2$ are nontrivial.

To shed the light on the topological phase transition when stacking different numbers of layers of V$_2$HfC$_2$(OH)$_2$, the (0001) surface states are calculated from double layers to quadruple layers. Their energy bands along $\Sigma$ and $\Lambda$-axes near the $\Gamma$ point are shown in FIG.~\ref{multilayers}. The distances between the neighboring layers are 20~{\AA} and the vacuum sizes of these supercells are 60~{\AA} as shown in the Supplemental Material (FIG. S12). In double and triple layers, the valence bands are separated by the direct gaps, and their energy bands are trivial. However, the energy bands become topologically nontrivial in quadruple layers and the valence band and conduction band are connected by a Dirac cone. In these calculations, the multilayers have the AA-stacking type, which means that each monolayer is translated vertically along the $c$-axis. To investigate the influence of the staking type to the topology of energy bands, the quadruple layers with ABAB stacking type, which means that the alternative B layer is translated by the in-plane vector $\vec{a}/3-\vec{b}/3$ with respect to A layer, is adopted. The sketch of the structures and the energy bands near the $\Gamma$-point are given in the Supplemental Material (FIG. S12). The computational results indicate that the stacking type has no significant influence on the topological properties of the energy bands because the Dirac cone is observed in the calculations.
\section{Conclusions}
\label{Conclusion}

We highlight the influence of the IP states on the topological properties in the OH-functionalized M$'_2$M$''$C$_2$(OH)$_2$, where M$'$ = V, Nb, Ta and M$''$ = Ti, Zr, Hf. On the basis of the first-principles calculations, an energy band inversion involving IP states is proposed. It is demonstrated that the topological properties can be modulated by the interlayer distances. IP states have spatial extensive wave functions and couple the neighboring layers. By tuning the interlayer distances from 30~{\AA} to 20~{\AA}, the hybridization between the IP states from the neighboring layers trigger the energy band inversion. The trivial materials become strong TIs. We hope that this new mechanism will facilitate the application of multilayers with low work function as the topological materials in the future.

\bigskip

\begin{acknowledgements}
We would like to express our sincere thanks to the crew of the Center for Computational Materials Science of the Institute for Materials Research, Tohoku University for their continuous support. 

\end{acknowledgements}

%\suppinfo
\newpage

\begin{flushleft}
{\large \textbf{Captions} }
\end{flushleft}

\begin{figure}[tbp]
\includegraphics[width=1\textwidth]{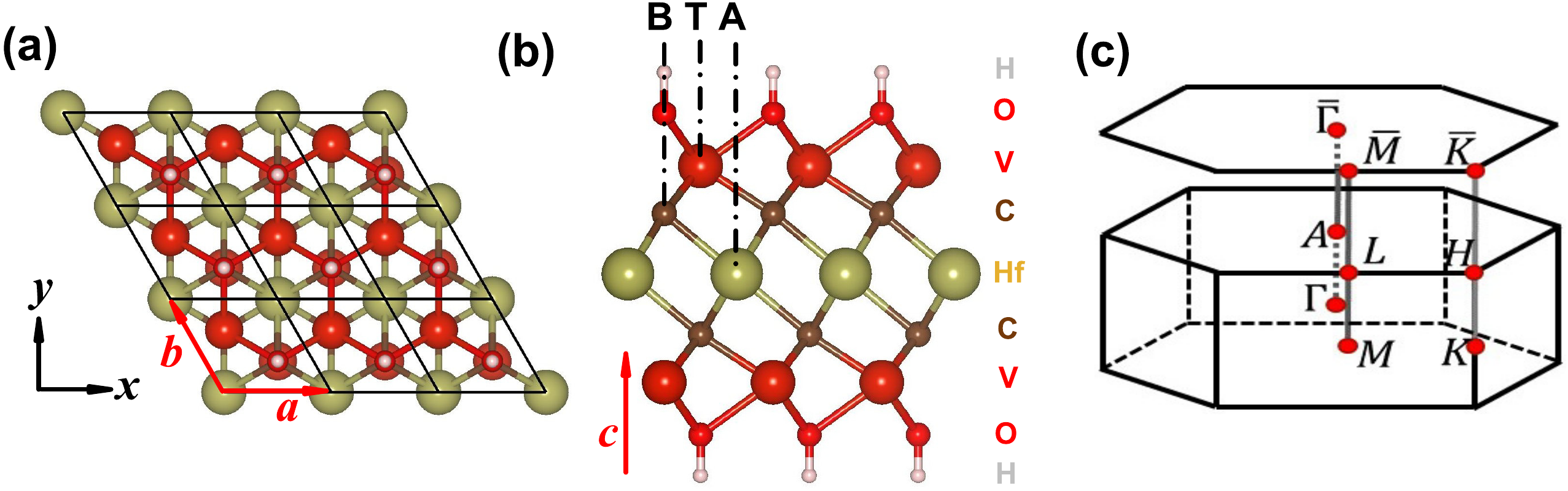}
\caption{M. Wang \emph{et. al.}} \label{structure_OH}
\end{figure}

\begin{figure}[tbp]
\includegraphics[width=1\textwidth]{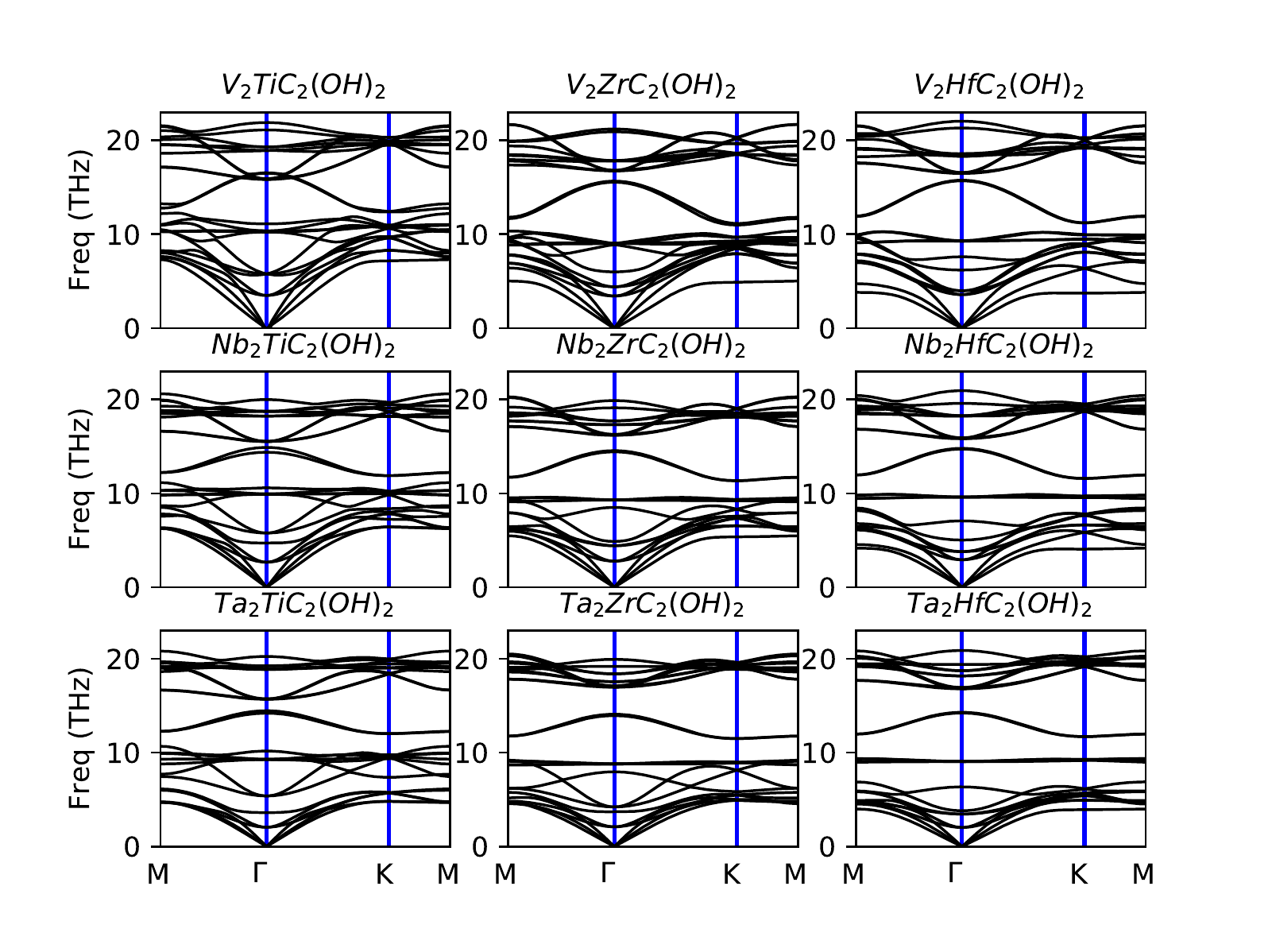}
\caption{M. Wang \emph{et. al.}} \label{phonon}
\end{figure}

\begin{figure}[tbp]
\includegraphics[width=1\textwidth]{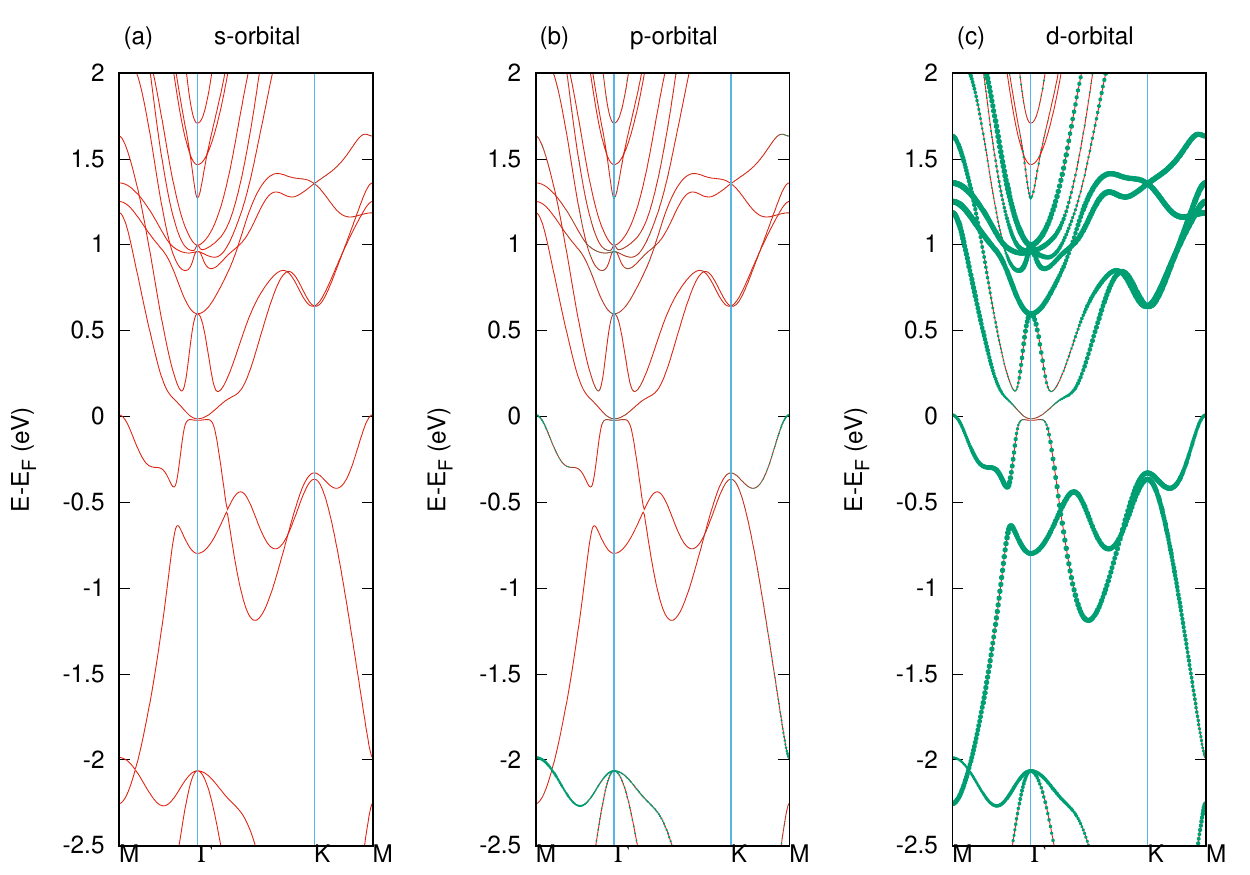}
\caption{M. Wang \emph{et. al.}} \label{fatband_V2Hf_30}
\end{figure}

\begin{figure}[tbp]
\includegraphics[width=1\textwidth]{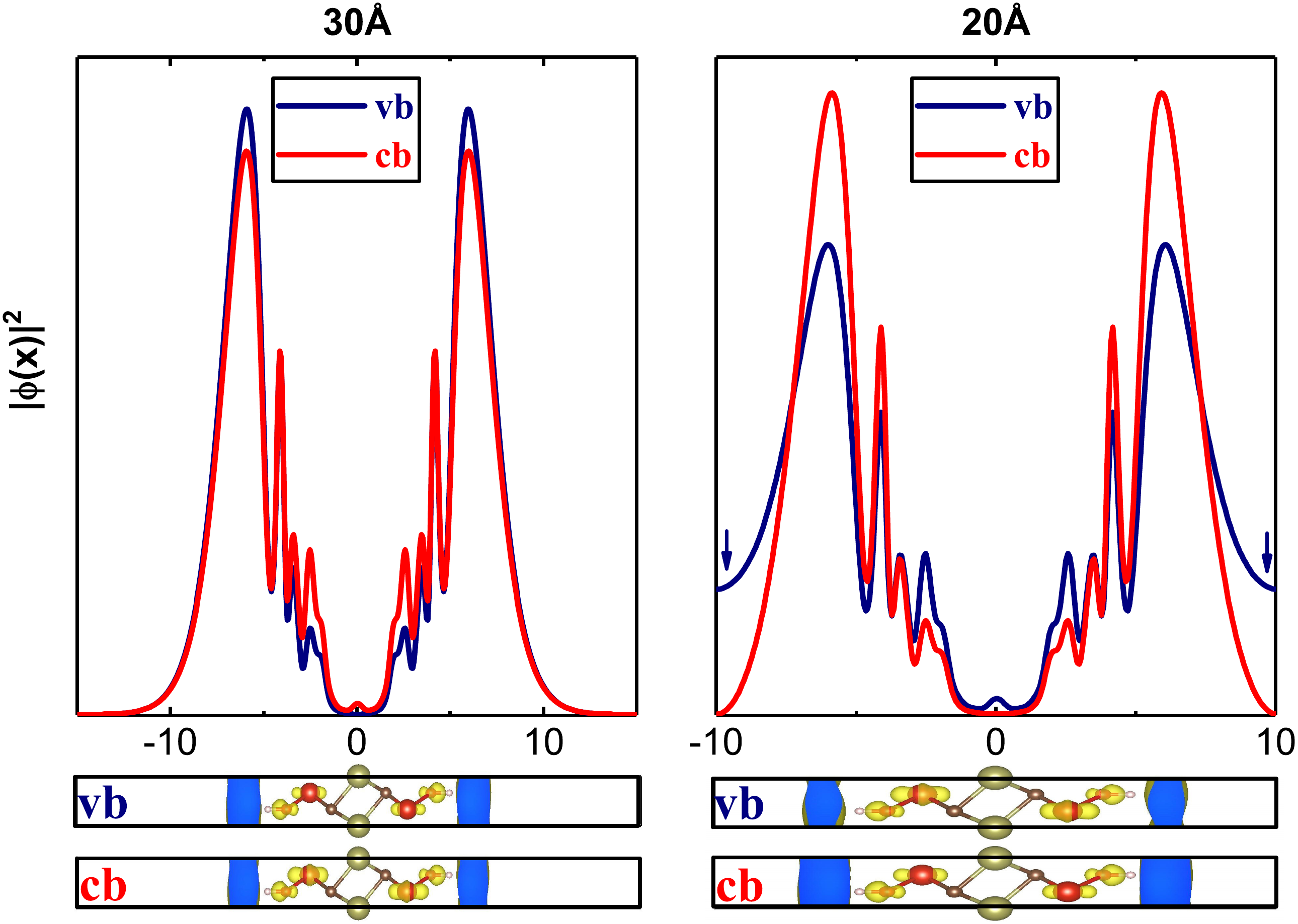}
\caption{M. Wang \emph{et. al.}} \label{chgden_V2Hf}
\end{figure}

\begin{figure}[tbp]
\includegraphics[width=1\textwidth]{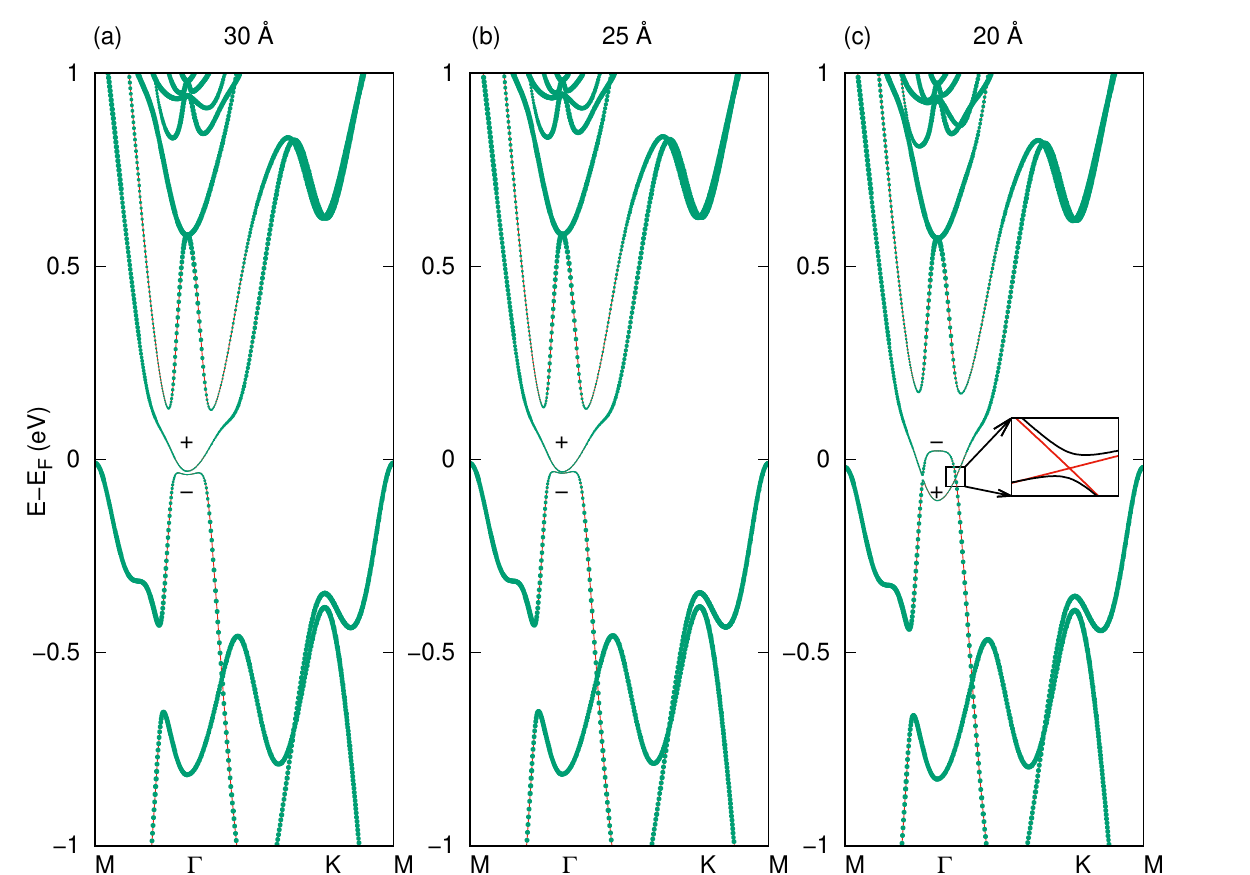}
\caption{M. Wang \emph{et. al.}} \label{band_distance}
\end{figure}

\begin{figure}[tbp]
\includegraphics[width=1\textwidth]{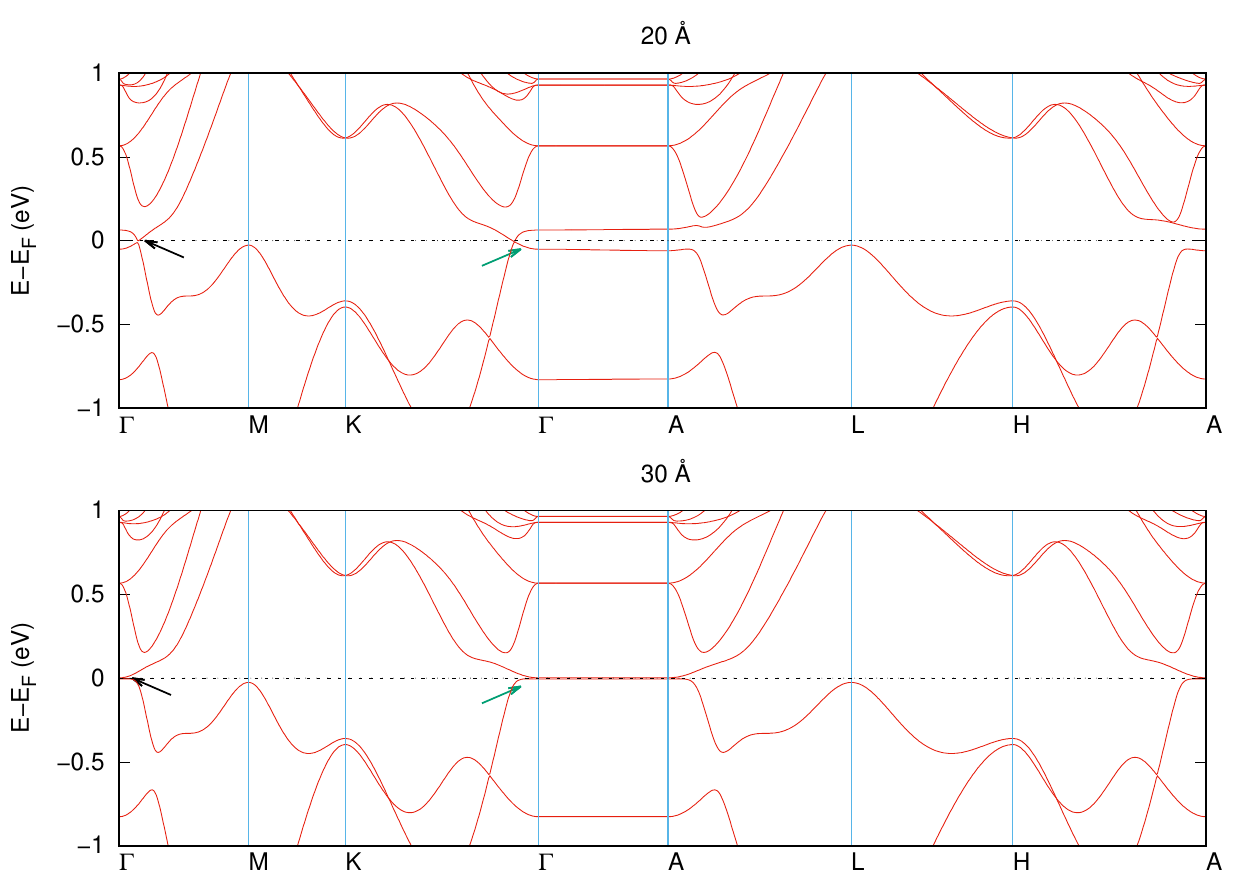}
\caption{M. Wang \emph{et. al.}} \label{3dband}
\end{figure}

\begin{figure}[tbp]
\includegraphics[width=1\textwidth]{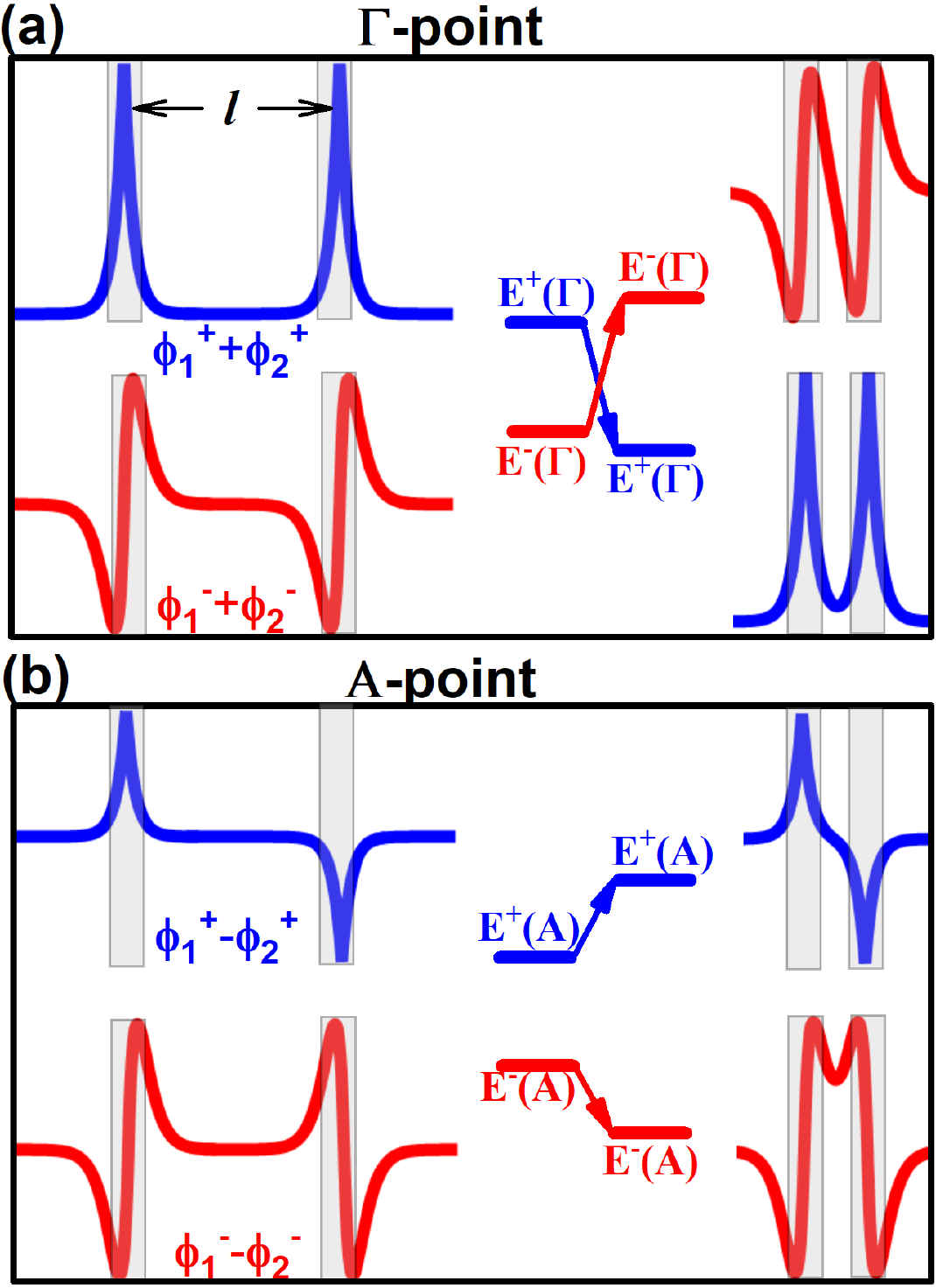}
\caption{M. Wang \emph{et. al.}} \label{band_inv}
\end{figure}

\begin{figure}[tbp]
\includegraphics[width=1\textwidth]{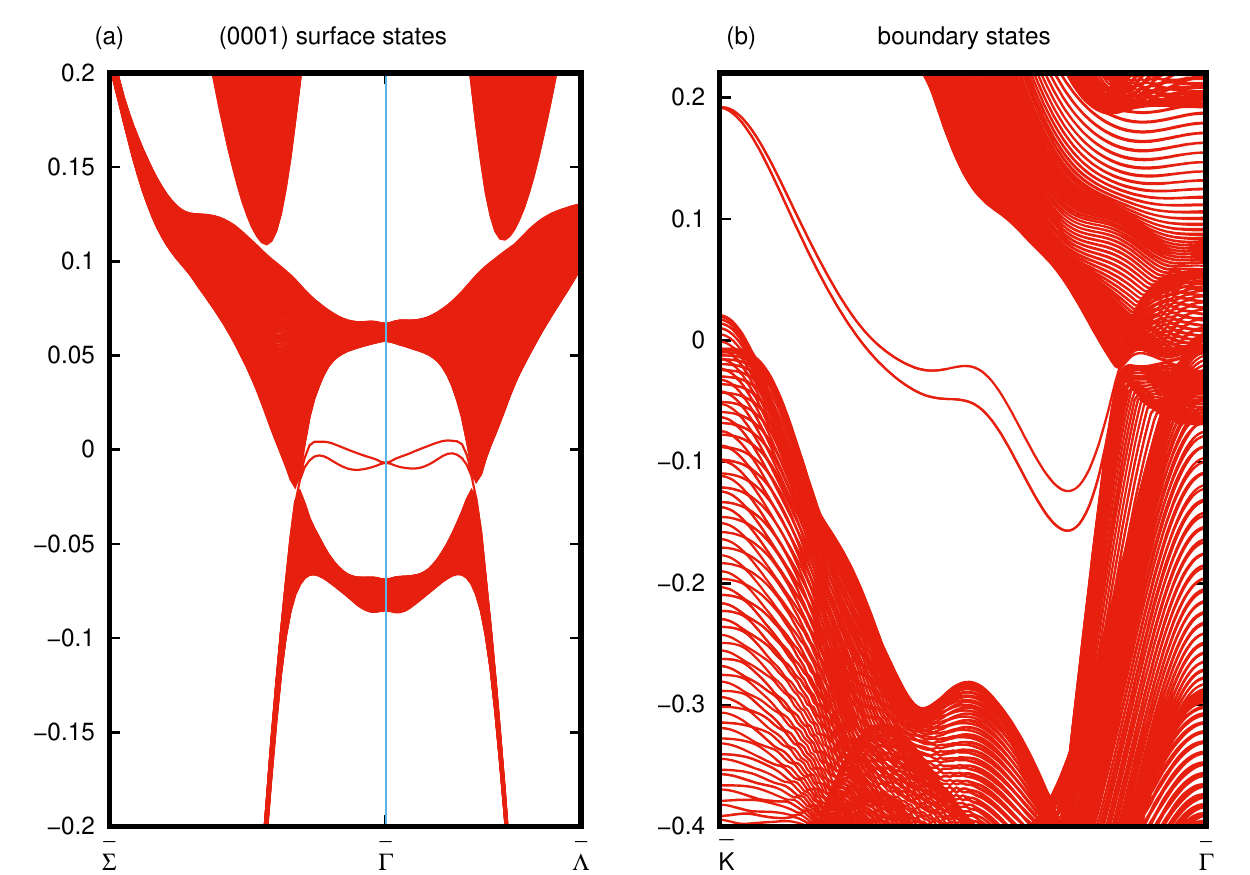}
\caption{M. Wang \emph{et. al.}} \label{edge_state}
\end{figure}

\begin{figure}[tbp]
\includegraphics[width=1\textwidth]{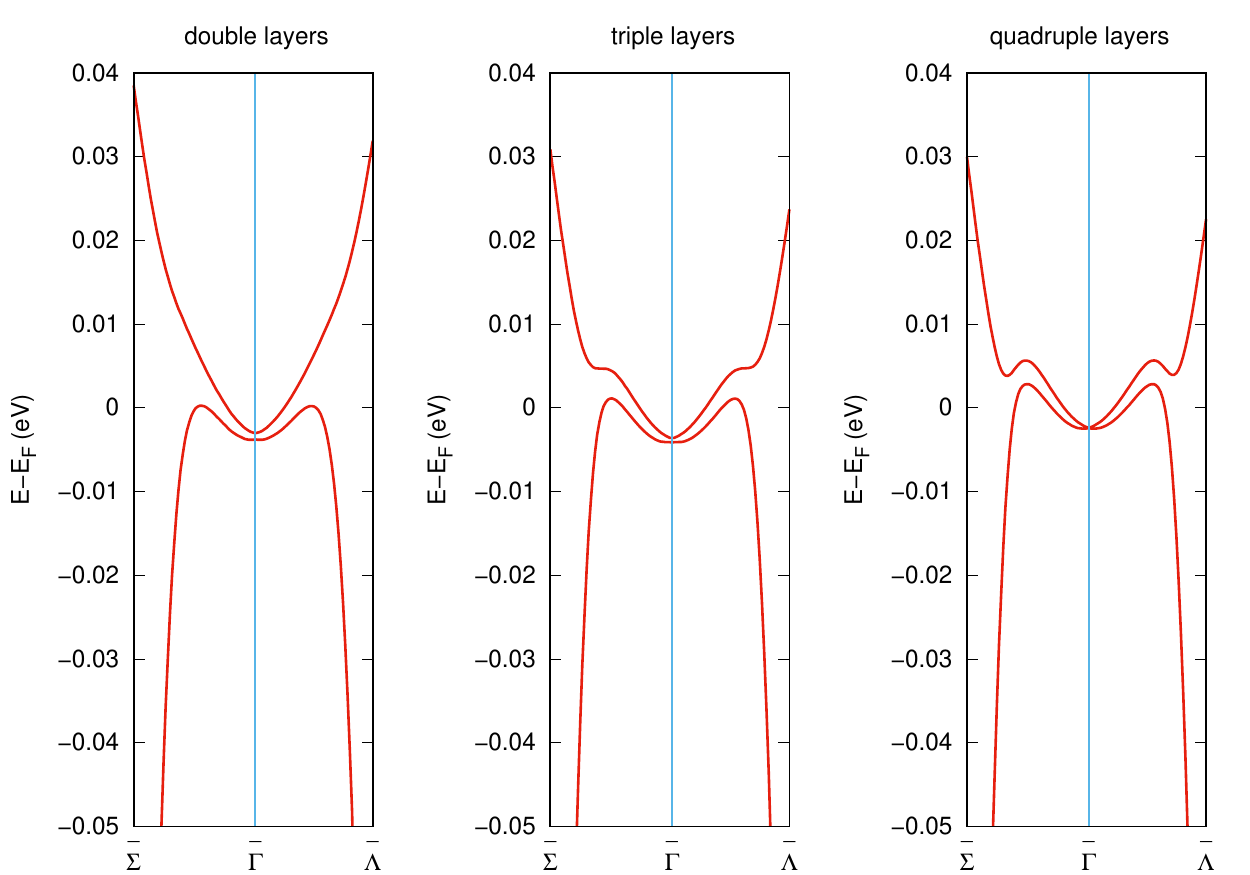}
\caption{M. Wang \emph{et. al.}} \label{multilayers}
\end{figure}

\begin{description}
\item{FIG. \ref{structure_OH}} The BB-structures of V$_2$HfC$_2$(OH)$_2$ monolayers with top (a) and side (b) views. The first Brillouin zone and its corresponding high symmetrical points are shown in panel (c). The box in (a) indicates the primitive cell.

\item{FIG. \ref{phonon}} The phonon spectra of M$'_2$M$''$C$_2$(OH)$_2$, where M$'$=V, Nb, Ta and M$''$=Ti, Zr, Hf.

\item{FIG. \ref{fatband_V2Hf_30}} The fat-bands of V$_2$HfC$_2$(OH)$_2$ with 30~{\AA} interlayer distance. The wave functions are projected onto s-orbitals (a), p-orbitals (b) and d-orbitals (c) of all atoms in the cell, respectively. The sizes of the dots are proportional to the contribution weights from these orbitals. 

\item{FIG. \ref{chgden_V2Hf}} The averaged decomposed charge densities of the valence band (vb) and the conduction band (cb) at $\Gamma$-point. They are averaged along the $c$-axis and the corresponding side views of the decomposed charge densities are below them. The interlayer distances in panels (a) and (b) are {30~\AA} and {20~\AA}, respectively. The isovalue of the charge density is 0.004.

\item{FIG. \ref{band_distance}} Panels (a)-(c) show the fat-bands of V$_2$HfC$_2$(OH)$_2$ with 30, 25 and 20~{\AA} interlayer distances, respectively. The wave functions are projected onto atomic orbitals in the cell. The sizes of the dots are proportional to the contribution weights from atomic orbitals. The inset shows the opening of the gap in the presence of SOC.

\item{FIG. \ref{3dband}} The energy bands along high symmetrical axes. The interlayer distances are 20~{\AA} in panel (a) and 30~{\AA} in panel (b), respectively.

\item{FIG. \ref{band_inv}} The energy level shifts at $\Gamma$ (a) and $A$-points (b) when the interlayer distances $l$ are reduced. $\phi_1^{\pm}$ and $\phi_2^{\pm}$ are IP states from the neighboring layers with different parities. The energy band inversion occurs at $\Gamma$-point.

\item{FIG. \ref{edge_state}} The calculated surface states of V$_2$HfC$_2$(OH)$_2$ multilayers. Panel (a) is (0001) surface state around $\Gamma$-point. Panel (b) is the boundary (cutting along $y$-axis) states. The interlayer distance $l$ is 20~{\AA} and the Fermi level is located at zero energy. 

\item{FIG. \ref{multilayers}} Panel (a) to panel (c) are the (0001) surface states of (a) double, (b) triple and (c) quadruple layers of V$_2$HfC$_2$(OH)$_2$.
\end{description}

\clearpage\newpage

\clearpage\newpage

\clearpage\newpage

\clearpage\newpage

\clearpage\newpage

\clearpage\newpage

\clearpage\newpage

\clearpage\newpage

\clearpage\newpage

\end{document}

\title{Supplemental Material for Topological Phase Transition Induced by Image Potential States in MXenes: A Theoretical Investigation}

\author{Mengying Wang}
\affiliation{Department of Physics, Shanghai Normal University, Shanghai 200234, China}

\author{Mohammad Khazaei}
\affiliation{Department of Physics, Yokohama National University, Yokohama 240-8501, Japan}

\author{Yoshiyuki Kawazoe}
\affiliation{New Industry Creation Hatchery Center, Tohoku University, Sendai, 980-8579, Japan}
\affiliation{School of Physics, Institute of Science and Center of Excellence in Advanced Functional Materials, Suranaree University of Technology, Nakhon Ratchasima 30000, Thailand}

\author{Yunye Liang}
\email{E-mail: liangyunye@shnu.edu.cn}
\affiliation{Department of Physics, Shanghai Normal University, Shanghai 200234, China}

\date{\today}% It is always \today, today,

\widetext

\maketitle

\begin{figure}[tbp]
\includegraphics[width=1\textwidth]{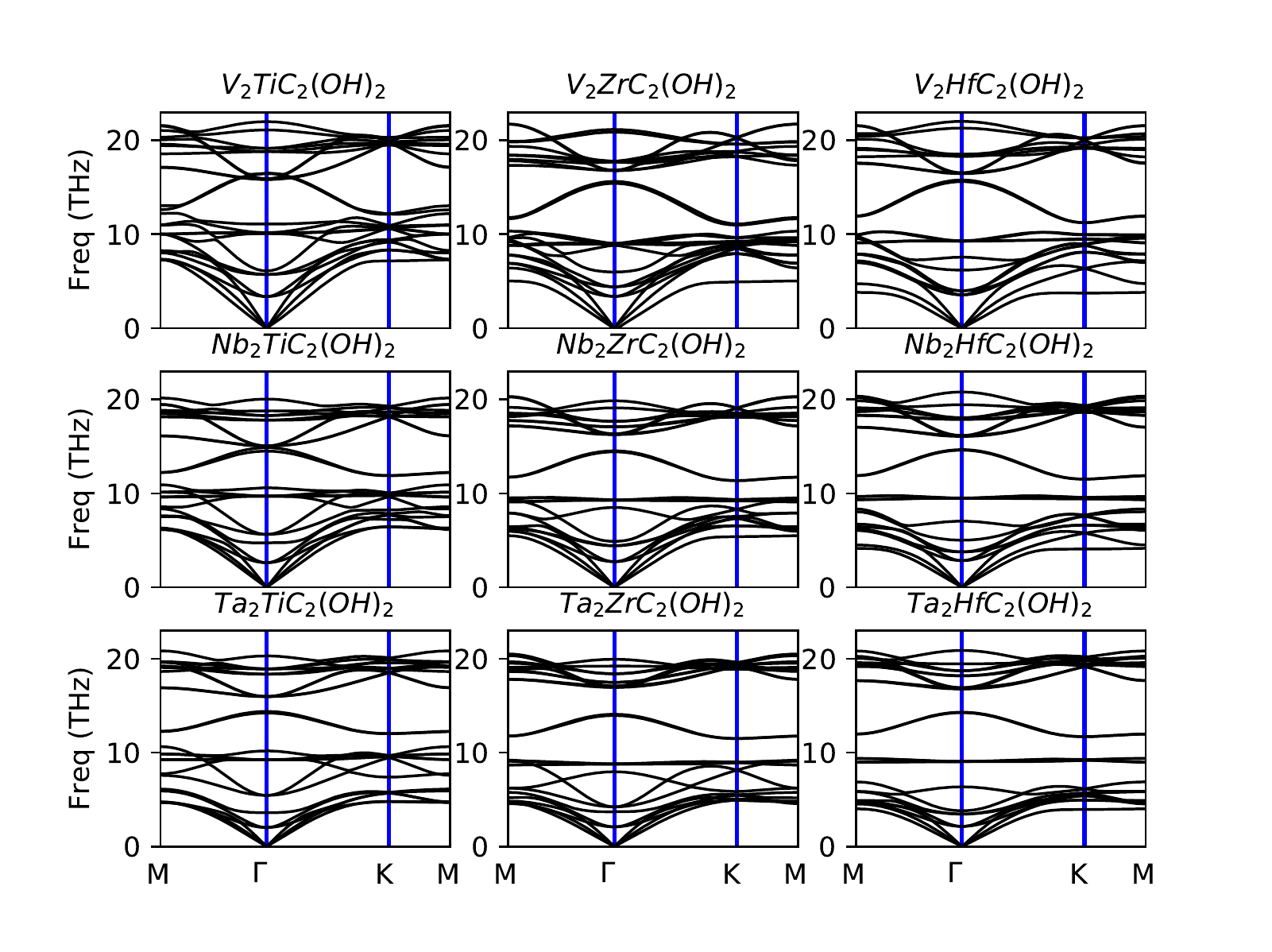}
\caption{M. Wang \emph{et. al.}} \label{phonon}
\end{figure}

\begin{figure}[tbp]
\includegraphics[width=1\textwidth]{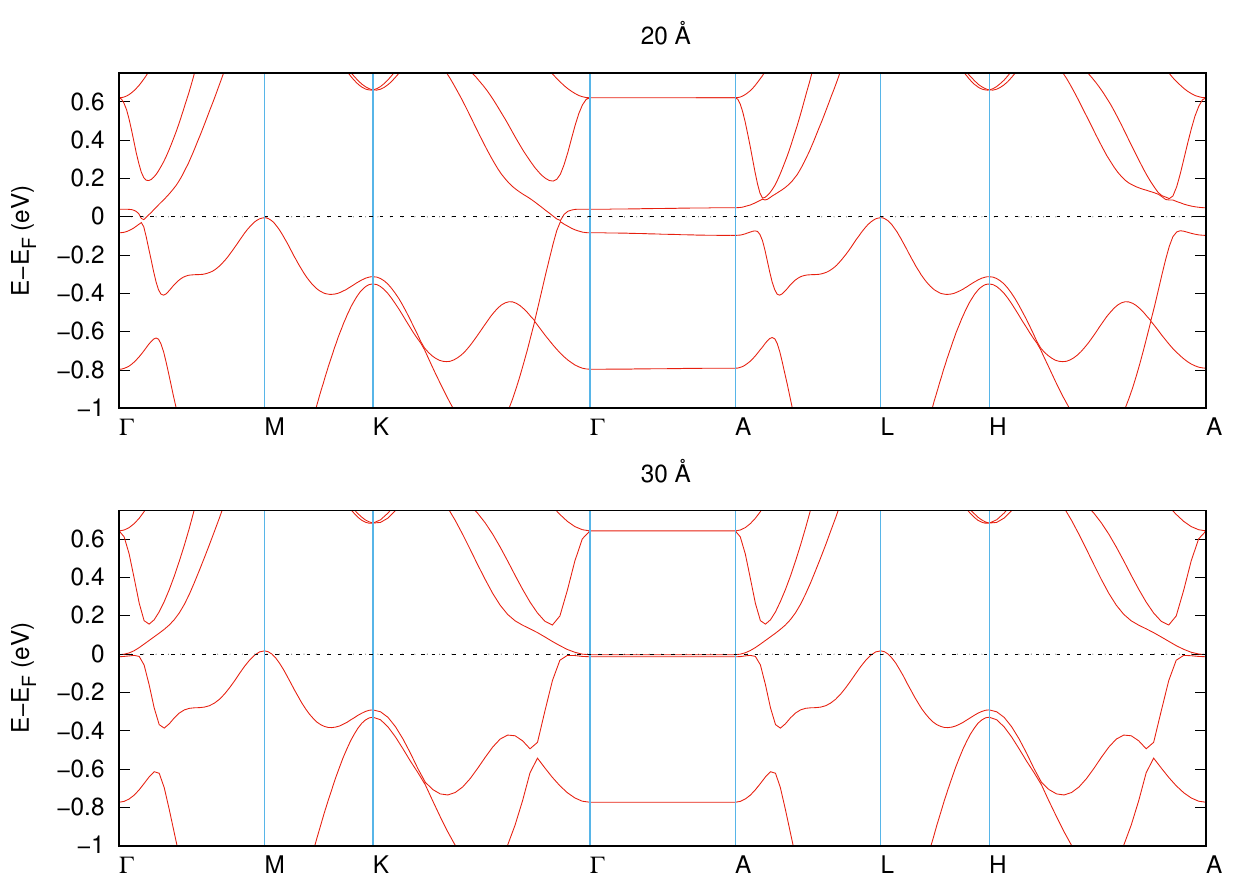}
\caption{M. Wang \emph{et. al.}} \label{vdW}
\end{figure}

\begin{figure}[tbp]
\includegraphics[width=1\textwidth]{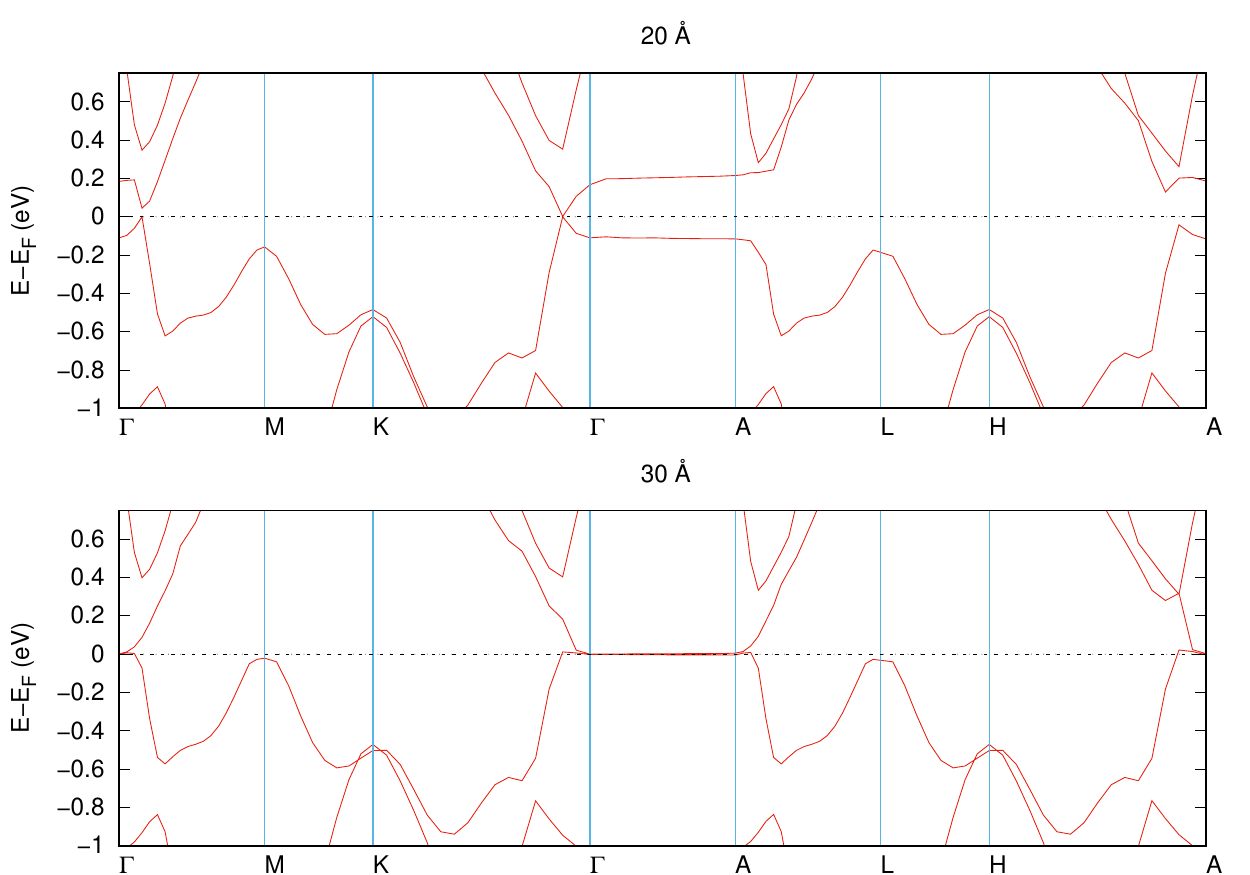}
\caption{M. Wang \emph{et. al.}} \label{hse}
\end{figure}

\begin{figure}[tbp]
\includegraphics[width=1\textwidth]{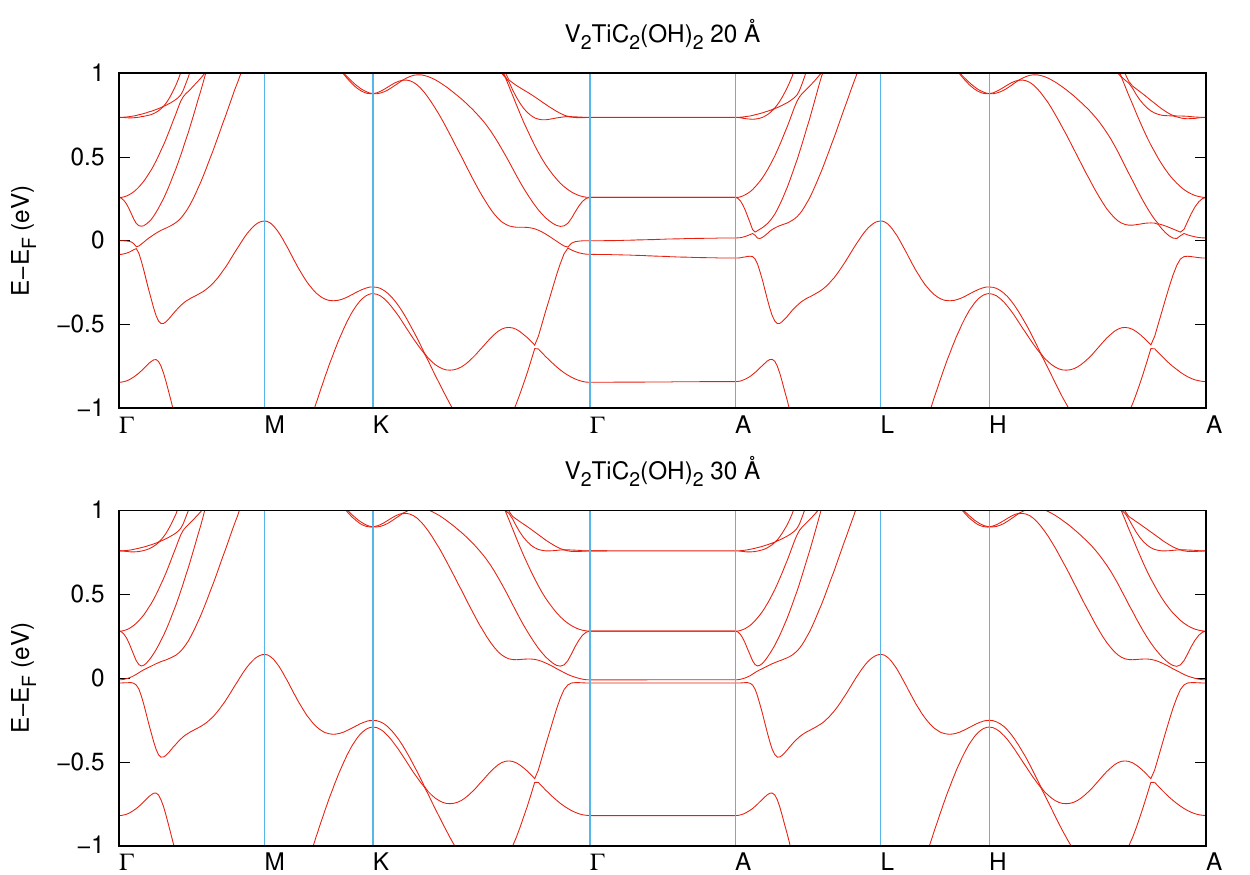}
\caption{M. Wang \emph{et. al.}} \label{V2TiC2O2H2}
\end{figure}

\begin{figure}[tbp]
\includegraphics[width=1\textwidth]{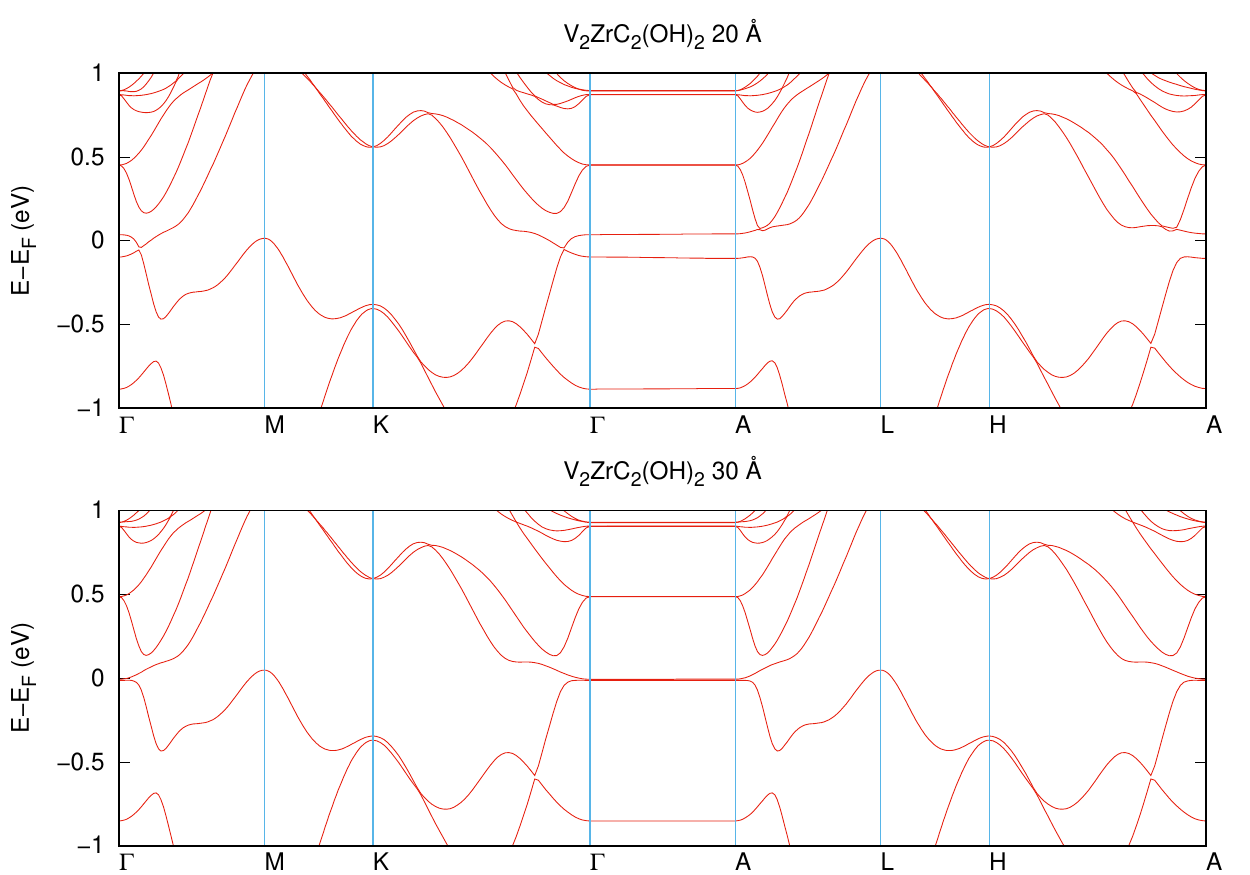}
\caption{M. Wang \emph{et. al.}} \label{V2ZrC2O2H2}
\end{figure}

\begin{figure}[tbp]
\includegraphics[width=1\textwidth]{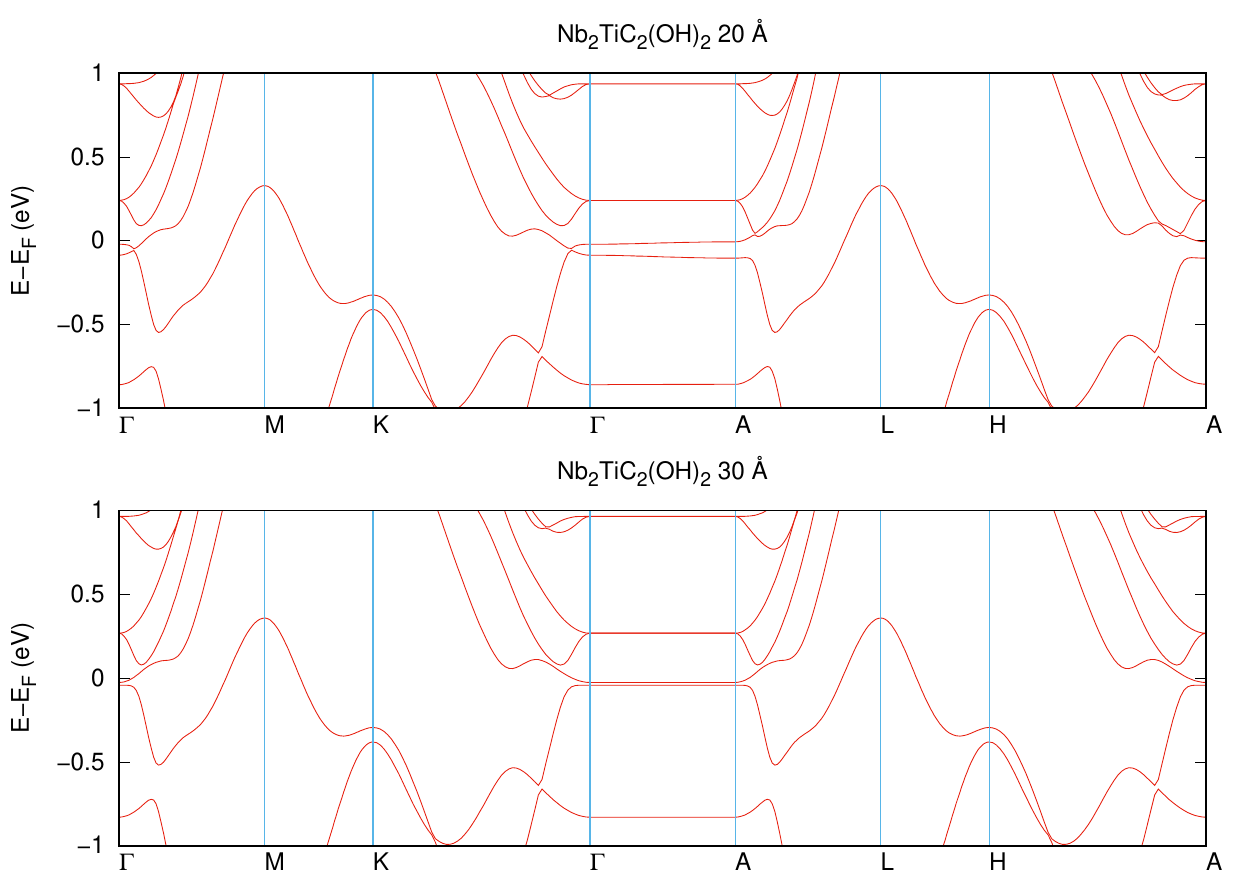}
\caption{M. Wang \emph{et. al.}} \label{Nb2TiC2O2H2}
\end{figure}

\begin{figure}[tbp]
\includegraphics[width=1\textwidth]{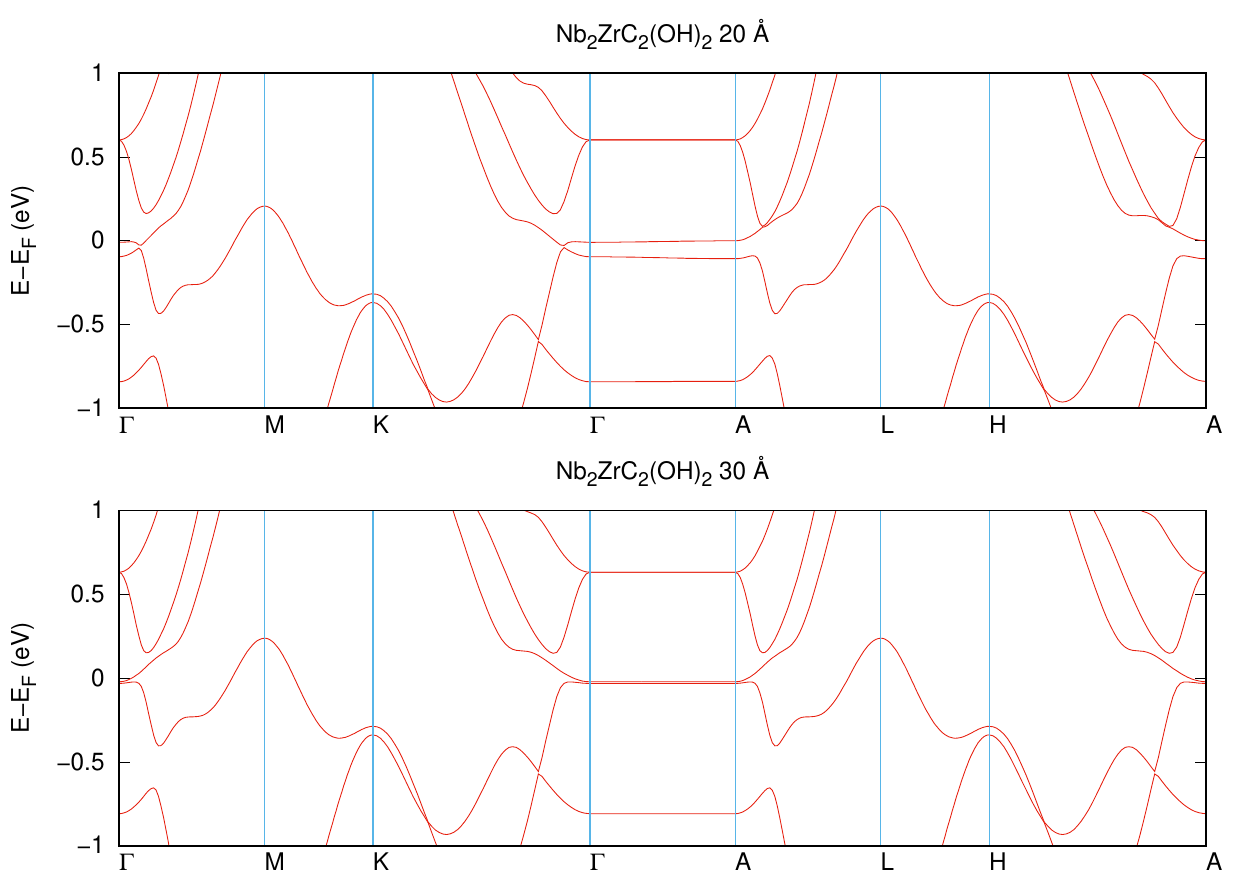}
\caption{M. Wang \emph{et. al.}} \label{Nb2ZrC2O2H2}
\end{figure}

\begin{figure}[tbp]
\includegraphics[width=1\textwidth]{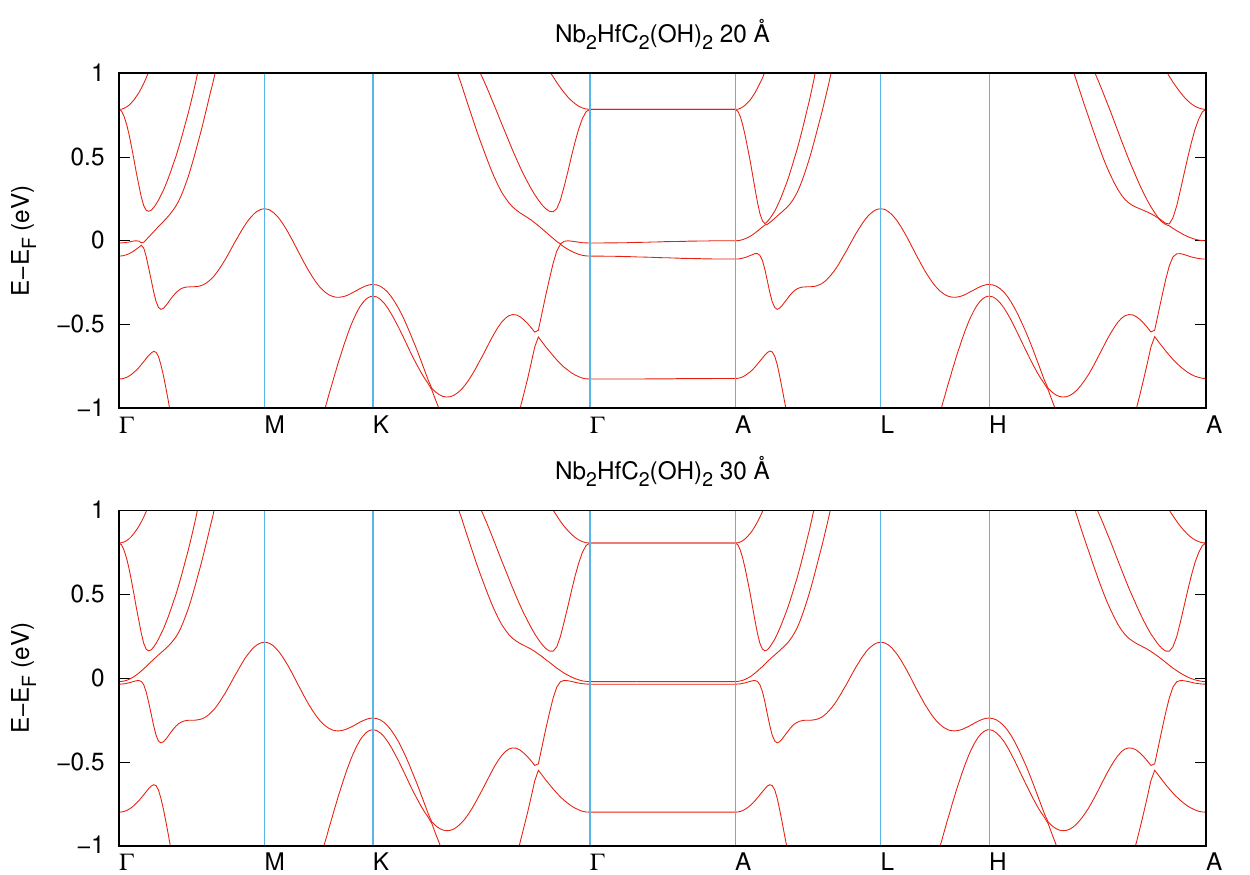}
\caption{M. Wang \emph{et. al.}} \label{Nb2HfC2O2H2}
\end{figure}

\begin{figure}[tbp]
\includegraphics[width=1\textwidth]{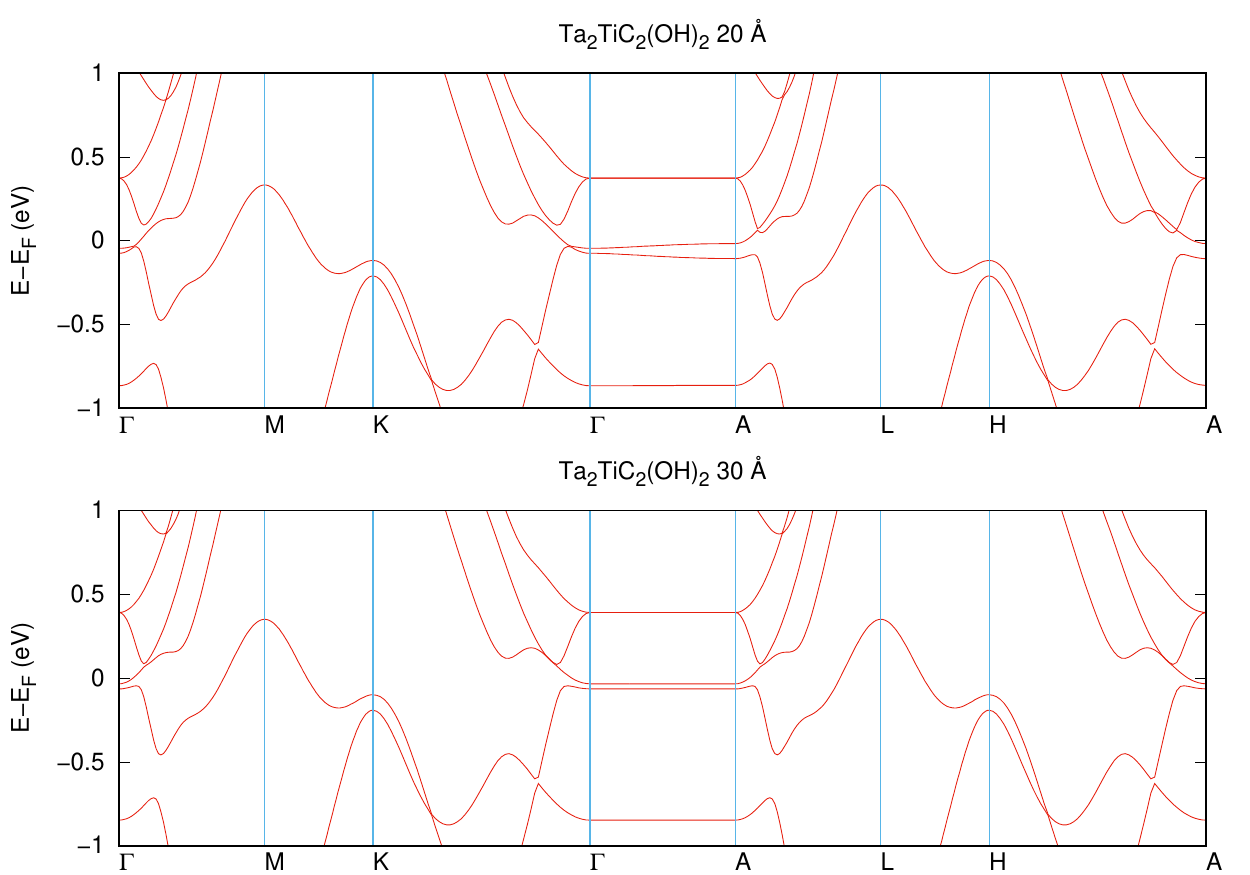}
\caption{M. Wang \emph{et. al.}} \label{Ta2TiC2O2H2}
\end{figure}

\begin{figure}[tbp]
\includegraphics[width=1\textwidth]{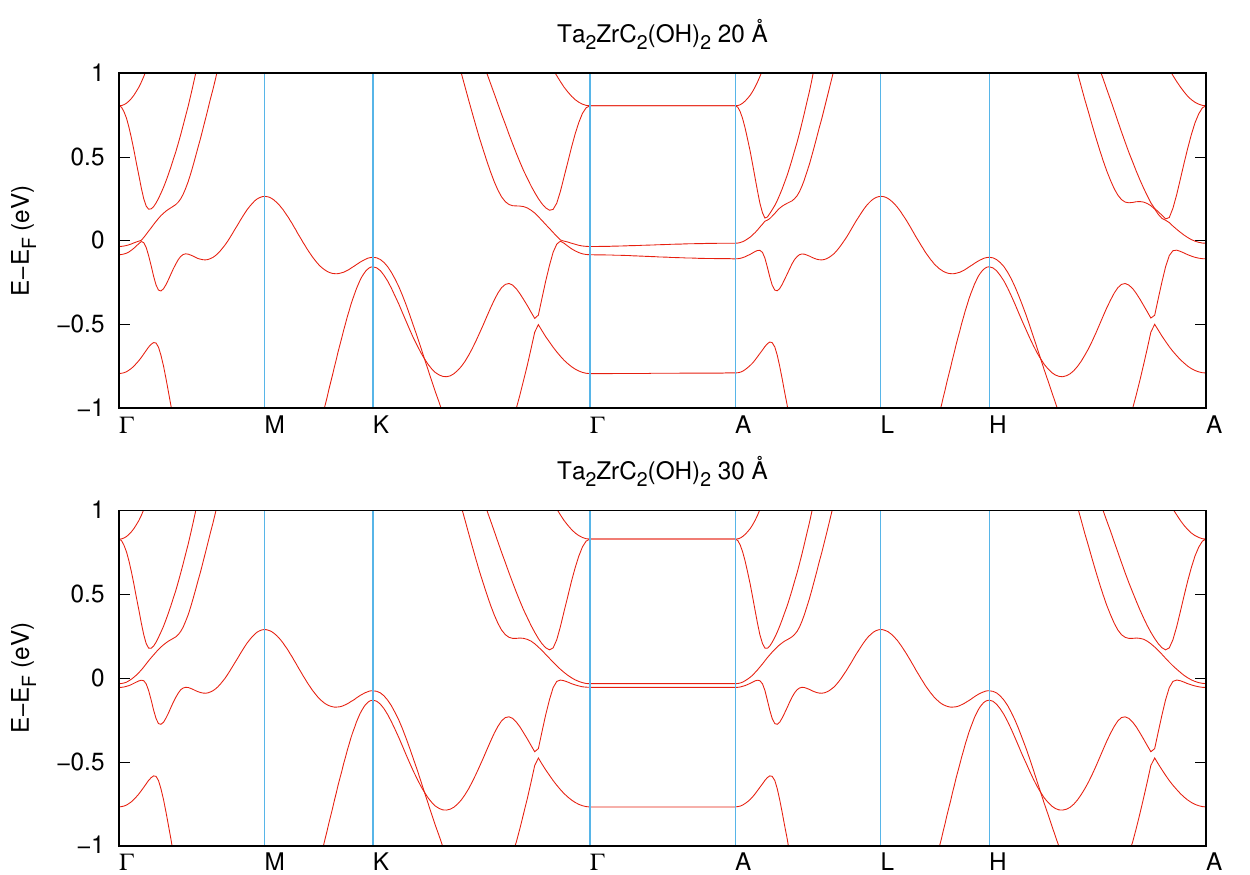}
\caption{M. Wang \emph{et. al.}} \label{Ta2ZrC2O2H2}
\end{figure}

\begin{figure}[tbp]
\includegraphics[width=1\textwidth]{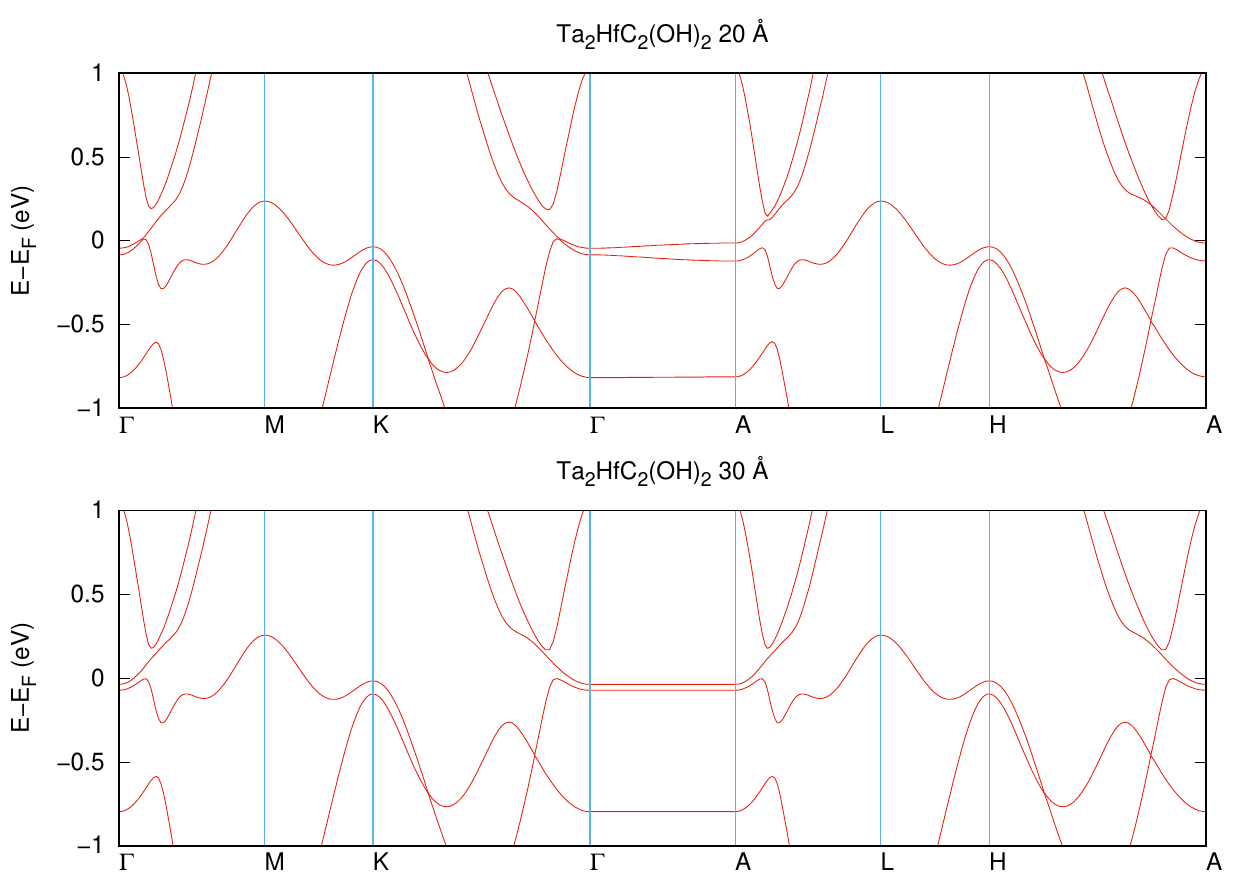}
\caption{M. Wang \emph{et. al.}} \label{Ta2HfC2O2H2}
\end{figure}

\begin{figure}[tbp]
\includegraphics[width=1\textwidth]{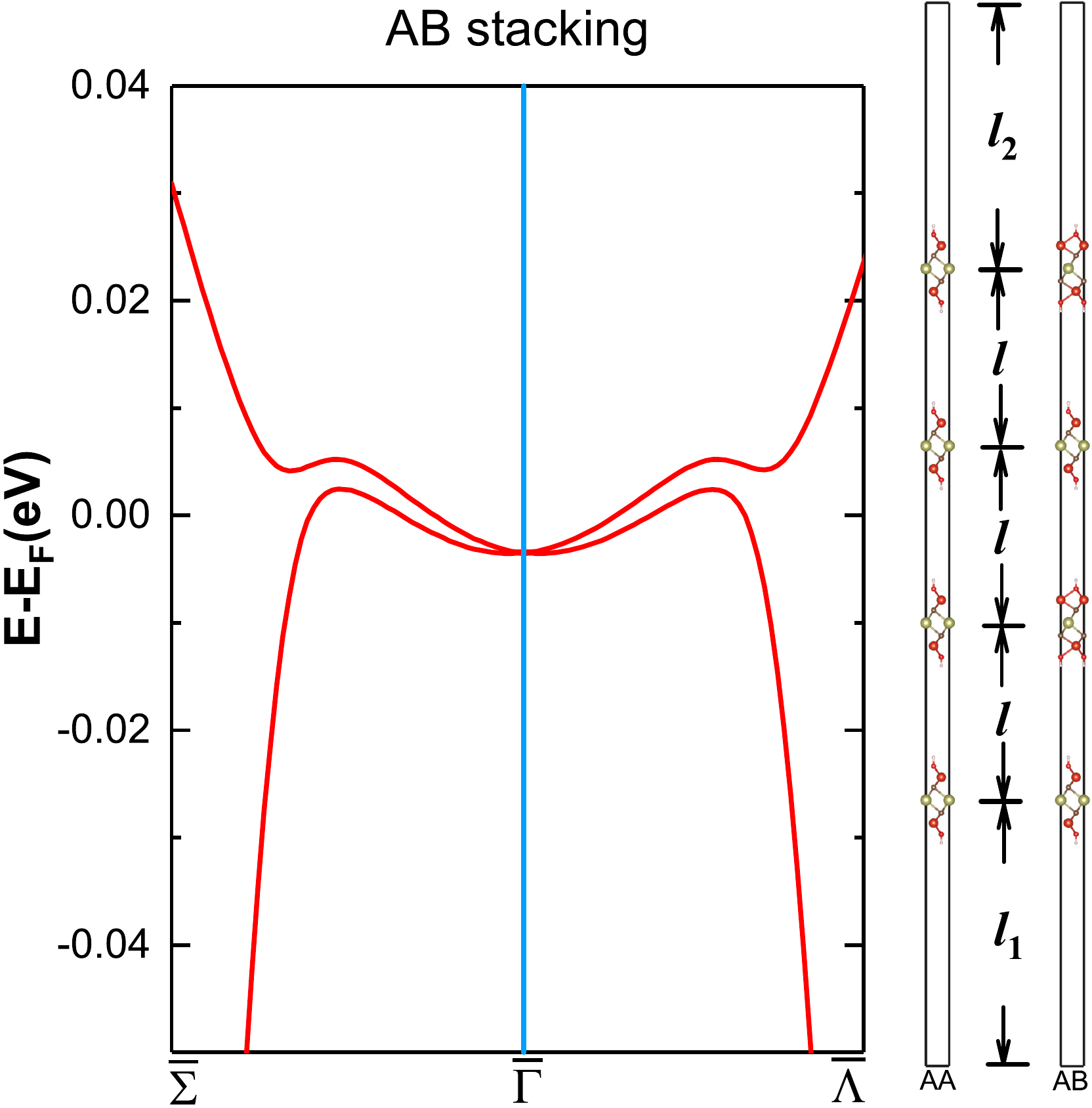}
\caption{M. Wang \emph{et. al.}} \label{stacking}
\end{figure}

\begin{description}
\item{FIG.~\ref{phonon}} The phonon spectra of M$'_2$M$''$C$_2$(OH)$_2$, where M$'$=V, Nb, Ta and M$''$=Ti, Zr, Hf. The vacuum sizes are 20~{\AA}.  The force constants are extracted from 4$\times$4$\times$1 supercells with 5$\times$5$\times$3 $k$-mesh. 

\item{FIG.~\ref{vdW}} The energy bands of V$_2$HfC$_2$(OH)$_2$ with the consideration of van der Waals (vdW) interaction. The vacuum sizes are 20~{\AA} (upper) and 30~{\AA} (lower). The SOC is not included.

\item{FIG.~\ref{hse}} The energy bands of V$_2$HfC$_2$(OH)$_2$ given by HSE06 functional. The vacuum sizes are 20~{\AA} (upper) and 30~{\AA} (lower). The SOC is not included.

\item{FIG.~\ref{V2TiC2O2H2}} The energy bands of V$_2$TiC$_2$(OH)$_2$ multilayers with different interlayers calculated from PBE functional. In the upper panel, the interlayer is 20{~\AA}, and in the lower panel, the interlayer is 30{~\AA}. The Fermi level is zero, and the SOC is not included.

\item{FIG.~\ref{V2ZrC2O2H2}} The energy bands of V$_2$ZrC$_2$(OH)$_2$ multilayers with different interlayers calculated from PBE functional. In the upper panel, the interlayer is 20{~\AA}, and in the lower panel, the interlayer is 30{~\AA}. The Fermi level is zero, and the SOC is not included.

\item{FIG.~\ref{Nb2TiC2O2H2}} The energy bands of Nb$_2$TiC$_2$(OH)$_2$ multilayers with different interlayers calculated from PBE functional. In the upper panel, the interlayer is 20{~\AA}, and in the lower panel, the interlayer is 30{~\AA}. The Fermi level is zero, and the SOC is not included.

\item{FIG.~\ref{Nb2ZrC2O2H2}} The energy bands of Nb$_2$ZrC$_2$(OH)$_2$ multilayers with different interlayers calculated from PBE functional. In the upper panel, the interlayer is 20{~\AA}, and in the lower panel, the interlayer is 30{~\AA}. The Fermi level is zero, and the SOC is not included.

\item{FIG.~\ref{Nb2HfC2O2H2}} The energy bands of Nb$_2$HfC$_2$(OH)$_2$ multilayers with different interlayers calculated from PBE functional. In the upper panel, the interlayer is 20{~\AA}, and in the lower panel, the interlayer is 30{~\AA}. The Fermi level is zero, and the SOC is not included.

\item{FIG.~\ref{Ta2TiC2O2H2}} The energy bands of Ta$_2$TiC$_2$(OH)$_2$ multilayers with different interlayers calculated from PBE functional. In the upper panel, the interlayer is 20{~\AA}, and in the lower panel, the interlayer is 30{~\AA}. The Fermi level is zero, and the SOC is not included.

\item{FIG.~\ref{Ta2ZrC2O2H2}} The energy bands of Ta$_2$ZrC$_2$(OH)$_2$ multilayers with different interlayers calculated from PBE functional. In the upper panel, the interlayer is 20{~\AA}, and in the lower panel, the interlayer is 30{~\AA}. The Fermi level is zero, and the SOC is not included.

\item{FIG.~\ref{Ta2HfC2O2H2}} The energy bands of Ta$_2$HfC$_2$(OH)$_2$ multilayers with different interlayers calculated from PBE functional. In the upper panel, the interlayer is 20{~\AA}, and in the lower panel, the interlayer is 30{~\AA}. The Fermi level is zero, and the SOC is not included.

\item{FIG.~\ref{stacking}} The surface state of V$_2$HfC$_2$(OH)$_2$ quadruple layers with ABAB stacking. The alternative B layer is translated by the in-plane vector $\vec{a}/3-\vec{b}/3$ with respect to A layer. The calculated surface states are protected topologically. The sketch of the AA and AB stacking structures are in the right side. $l$(=20~\AA) is the interlayer distance, and $l_1+l_2$ is the vacuum size of the supercell with $l_1=l_2=$30~\AA.
\end{description}

\newpage
\clearpage\newpage

\newpage
\clearpage\newpage

\newpage
\clearpage\newpage

\newpage
\clearpage\newpage

\newpage
\clearpage\newpage

\newpage
\clearpage\newpage

\newpage
\clearpage\newpage

\newpage
\clearpage\newpage

\newpage
\clearpage\newpage

\newpage
\clearpage\newpage

\newpage
\clearpage\newpage

\newpage
\clearpage\newpage
